# Interplay of Sensor Quantity, Placement and System Dimensionality on Energy Sparse Reconstruction of Fluid Flows

**Chen Lu and Balaji Jayaraman***

School of Mechanical and Aerospace Engineering, Oklahoma State University, Stillwater, OK, USA

E-mail: `balaji.jayaraman@okstate.edu`

21 June 2018

**Abstract.** Reconstruction of fine-scale information from sparse data is relevant to many practical fluid dynamic applications where the sensing is typically sparse. Fluid flows in an ideal sense are manifestations of nonlinear multiscale PDE dynamical systems with inherent scale separation that impact the system dimensionality. There is a common need to analyze the data from flow measurements or high-fidelity computations for stability characteristics, identification of coherent structures and develop evolutionary models for real-time data-driven control. Given that sparse reconstruction is inherently an ill-posed problem, the most successful approaches require the knowledge of the underlying basis space spanning the manifold in which the system resides. In this study, we employ an approach that learns basis from singular value decomposition (SVD) of training data to reconstruct sparsely sensed information at randomly sampled locations. This allows us to leverage energy sparsity with $l_2$ minimization instead of the more expensive, sparsity promoting $l_1$ minimization. Further, for unknown flow systems where only global operating parameters such as Reynolds ($Re$) number and raw data are available, it is often not clear what the optimal number of sensors and their placement for near-exact reconstruction needs to be. In this effort, we explore the interplay of data sparsity, sparsity of the underlying flow system and sensor placement on energy sparse reconstruction performance enabled by data-driven SVD basis. To this end, we investigate sparse convolution-based reconstruction performance by characterizing operational bounds for canonical laminar cylinder wake flows in both limit-cycle and transient regimes.

*Keywords*: sparse reconstruction, sensors, cylinder flow, SVD, POD, compressive sensing

## 1. Introduction

Unsteady and multiscale fluid flow phenomena are ubiquitous in engineering and geophysical settings and are often the focus of scientific investigation. Depending on the situation, one encounters either a data-sparse or a data-rich problem. In the data-sparse case, the goal is to recover more information about the dynamical system while in the data-surplus case, the goal is to reduce the information into a simpler form for analysis or evolutionary prediction and then recover the high-dimensional state. For many practical fluid flow applications, accurate simulations may not be feasible for a multitude of reasons including, lack of accurate models, unknown governing equations or extremely complex boundary conditions.

In such situations, measurement data represents the absolute truth and is often acquired from very few probes that limits potential for in-depth analysis. A common recourse is to combine such sparse measurement with underlying knowledge of the flow system, either in the form of idealized simulations or phenomenology or knowledge of a sparse basis to recover detailed information. The former approach is termed as data assimilation while we refer to the latter as Sparse Convolution-based Reconstruction (SCR) or commonly Sparse Reconstruction (SR). In the absence of such a mechanism, the only method to identify structure information of the flow is to use phenomenology such as Taylor's frozen eddy hypothesis. On the other hand, simulations typically represent a data surplus setting. With growth in computing power, they often generate *big data* contributing to an ever growing demand for analytics and machine learning tools.

In spite of this added complexity, they offer the best avenue for analysis of realistic flows as one can identify and visualize coherent structures, perform well converged statistical analysis including quantification of spatio-temporal coherence and scale content. All of this is possible due to the high density of data probes in the form of computational grid points. At the same time, there is a need for simplifying these data into models with reduced dimensionality for practical, real-time analysis and decision making. Machine learning tools [1] such as data-driven Proper Orthogonal Decomposition (POD) [2, 3, 4] or other methods of projection onto a low-dimensional feature space [5] satisfy this need. However, once prediction happens in the feature space, the information needs to be mapped back to the high-dimensional state through a Sparse Convolution-based Reconstruction (SCR) step. Thus, tools for *encoding* information into a low-dimensional feature space (convolution) complement sparse reconstruction tools that help *decode* compressed information (deconvolution). This in essence is a key aspect of leveraging machine learning for fluid flow analysis [6, 7]. The other aspects of machine learning-driven study of fluid flows include building data-driven predictive models [5, 8, 9], pattern detection, and classification. In this article, we focus on the *decoding* problem of reconstructing high resolution fields data in both data-sparse and data-rich environments.

This work is primarily motivated by practical problems of flow sensing in the field or a laboratory. The rich physics of near-surface atmospheric flows is severely under sampled using just meteorological towers, even if designed carefully as part of a network such as the MESONET [10]. The entry of unmanned aerial systems (UAS) as mobile sensors has revolutionized geophysical sensing as the fast flying unmanned aerial vehicles (UAVs) can accrue many measurements over a characteristic turbulent large-eddy time-scale. This has significantly increased the quantity of field data available for learning the structure of micro-scale geophysical flows of relevance to pollutant/scalar transport and spreading of wild fires. However, such data is still severely under-resolved due to unavoidable limitations of scale in atmospheric sensing. As a result, the collected data is inherently sparse, and reconstruction of fine-scale information is needed to decipher the dominant coherent structures. Another example is flow sensing in the laboratory using Particle Image Velocimetry(PIV) [11, 12] which provides spatially resolved sensing, but can be unreliable for high fidelity analysis due to insufficient illumination, shadowing, obstructed view, and low seed density [13]. As a consequence, the resulting data is sparse, under-resolved, gappy, and almost always noisy which requires repairing the data using sparse reconstruction. PIV data can also be used for real-time control of turbulent flows by modulating specific coherent structures. This typically requires high-frequency and high-resolution sensing that makes real-time processing of the data expensive and challenging. The traditional approach is to acquire data first and then reduce it for use by the controller. Advances in compressive sensing (CS) [14, 15, 16, 17] and machine learning (ML) have opened the possibility of direct compressive sampling [6] to learn the relevant underlying structures from data. Thus, sparse data-driven decoding and reconstruction using sparse convolution ideas have been gaining popularity in recent years through their various manifestations such as Gappy Proper Orthogonal Decomposition (GPOD) [18, 19], Compressive Sensing (CS) [14, 15, 16, 17] and Kriging [20, 21, 13]. The overwhelming corpus of literature on this topic focus on theoretical expositions of the framework and demonstrations of performance. In this article, we investigate the interplay between system dimensionality, sensor quantity and placement on reconstruction quality for nonlinear fluid flows predicted by the numerical solution of partial differential equations (PDEs).

Sparse reconstruction is inherently ill-posed and underdetermined inverse problem where the number of constraints (i.e., sensor quantity) are much less than the number of unknowns (i.e., high resolution field). However, if the underlying system is sparse in a feature space then the probability of recovering a unique solution increases by solving the reconstruction problem in a lower-dimensional space. The core theoretical develoments of such ideas and their first practical applications happened in the realm of image compression and restoration [22, 16]. Data reconstruction techniques based on Karhunen-Loeve (K-L) procedure with $l_2$ minimization, also known as GPOD [23, 18, 19], was originally developed in the nineties to recover marred faces [23] in images. The fundamental idea is to utilize the POD basis computed offline from the data ensemble to *encode* the reconstruction problem into a low-dimensional feature space. This way, the sparse data can be used to recover the sparse unknowns in the feature space (i.e., sparse POD coefficients) by minimizing the $l_2$ errors. If the POD basis are not known *a priori*, an iterative formulation [18, 23] to successively approximate the POD basis and the coefficients was proposed. While this approach has been shown to work in principle [18, 20, 24], it is prone to numerical instabilities and inefficiency.

Advancements in the form of a progressive iterative reconstruction framework [20] are effective, but impractical for real-time application. A major issue with POD-basis is that they are data-driven and hence cannot be generalized, but are optimally sparse for the given data. This requires that they be generated offline and be used for efficient online sparse reconstruction using little sensor data. However, if training data are unavailable or if the prediction regime is not spanned by the precomputed basis, then the econstruction becomes *untenable*.

A way to overcome the above limitations is to use generic basis such as wavelets [25] or Fourier-based kernels. Such choices are based on the assumption that most systems are sparse in the feature spaces. This is particularly true for image processing applications but may not be optimal for fluid flows whose dynamics obey the PDEs. While avoiding the cost of computing the basis offline, such approaches run into sparsity issues as the basis do not optimally encode the underlying dynamical system. Thus, once again the reconstruction problem can be ill-posed even when solving in the feature space because the number of sensors could be smaller than the system dimensionality for the choice of basis. $l_2$ minimization produces a solution with sparsity mathcing the dimensionality of the feature space, thus requiring sensor quantity exceeding the system dimensionality. The magic of Compressive Sensing (CS) [14, 15, 16, 17] is in its ability to overcome this constraint by seeking a solution that can be less sparse than the dimensionality of the feature space using $l_1$-minimized norm reconstruction. Such methods have been successfully applied in image processing using Fourier or wavelet basis and also to fundamental fluid flows [9, 7, 6, 26, 27, 28]. Compressive sensing essentially looks for a sparse solution through the $l_1$ norm minimization of the sparse coefficients by solving a convex optimization problem that is computationally tractable and thereby, avoid the tendency of $l_2$-based methods to overfit the data. In recent years, compressive sensing-type $l_1$ reconstruction using POD basis has been employed successfully for reconstruction of sparse PIV data [6] and pressure measurements around a cylinder surface [7]. Since the POD basis are data-driven, they represent an optimal basis for reconstruction and require the least quantity of sensor measurements for a given reconstruction quality. However, the downside is the requirement of highly sampled training data as a one-time cost to build a library of POD bases. Such a framework has been attempted in [7] where POD modes from simulations over a range of Reynolds (*Re*) numbers of a cylinder wake flow were used to populate a library of bases and then used to classify the flow regime based on sparse measurements. In order to reduce this cost, one could also downsample the measurement data and learn the POD bases as per [6]. Recent efforts also combine CS with data-driven predictive ML tools such as Dynamic Mode Decomposition (DMD) [29, 30] to identify flow characteristics and classify into different stability regimes [28]. In the above, SR is embedded into the analysis framework for extracting relevant dynamical information.

Both SCR and CS can be viewed as generalizations of sparse regression in a higher-dimensional basis space. This way, one can relate SCR to other statistical estimation methods such as Kriging. Here, the data is represented as a realization of a random process that is stationary in the first and second moments. This allows one to interpolate information from the known to unknown data locations by employing a kernel (Commonly Gaussian) in the form of a variogram model and the weights are learned under conditions of zero bias and minimal variance. The use of kriging to recover flow-field information from sparse PIV data has been reported [20, 21, 13] with encouraging results.

The underlying concept in all the above described techniques is that they solve the reconstruction inverse problem in a feature or basis space where the number of unknowns are comparable to the number of constraints (sparse sensors). This mapping is done through a convolution or projection operator that can be constructed from data or kernel functions. Hence the reason we refer to this class of methods as sparse convolution-based reconstruction (SCR) in the same vein as sparse convolution-based Markov models [5, 9, 31, 32]. This requires existence of an optimal sparse basis space in which the physics can be represented. This exists in the form of wavelets and Fourier functions for many common applications in image and signal processing, but may not be optimally sparse for fluid flow solutions to PDEs. Hence, the reason data-driven basis such as POD/PCA [3, 4] are popular. Further, since they are optimally sparse, such methods can reconstruct with very little amount of data as compared to say, Kriging that employ generic Gaussian kernels. In this article, we aim to answer the following questions that will help understand the performance limits of these techniques for complex physical phenomena: (i) What is the optimal sparsity level of the solution to be reconstructed? (ii) How does the optimal sparsity levels for reconstruction depend on the choice of basis? (iii) What is the role of sensor locations on the reconstruction errors? Answering these questions will allow for such techniques to be employed effectively in practice. However, this is a challenging undertaking as for most practical situations one has access only to sparse data and the dynamics of the underlying system is unknown.

## 2. Formulating the Sparse Reconstruction Problem

Given a high resolution data representing the state of the flow system at any given instant denoted by $X \in \mathbb{R}^N$, its corresponding sparse representation given by $\tilde{X} \in \mathbb{R}^P$ with $P \ll N$. Then, the sparse reconstruction problem is to recover $X$, when given $\tilde{X}$ along with information of the sensor locations in the form the measurement matrix $C$ as shown in equation 1. The measurement matrix $C$ determines how the sparse data $\tilde{X}$ is collected from $X$. Variables $P$ and $N$ are the number of sparse measurements and the dimension of the high resolution field, respectively.

$$\tilde{X} = CX. \tag{1}$$

In this article, we focus on vectors $X$ that have a sparse representation in a basis space $\mathbf{\Phi} \in \mathbb{R}^{N \times K}$ such that $K \ll N$ and yielding $X = \mathbf{\Phi}a$. Naturally, when one loses the information about the system, the recovery of said information is not absolute as the reconstruction problem is ill-posed, i.e., more unknowns than equations in equation 1.

Thus the most straightforward approach to recover the solution $X$ by computing the inverse of $C$ as shown in Equation 2 is *not possible*.

$$C^{-1}\tilde{X} = X. \tag{2}$$

## 2.1. Sparse Reconstruction Theory

Sparse reconstruction theory has strong connections to the field of inverse problems and hence finds mention directly or indirectly in diverse fields of study such as a geophysics [33, 34], image processing [35, 36] and broadly speaking, inverse problems [37]. In this section, we formulate the reconstruction problem which has been presented in CS literature [14, 16, 38, 39, 40]. Many signals tend to be "compressible", i.e., they are sparse in some $K$-sparse basis $\boldsymbol{\Phi}$ as show below:

$$X = \sum_{i=1}^{N_b} \phi_i a_i \quad \text{or} \quad X = \boldsymbol{\Phi} a, \tag{3}$$

where $\boldsymbol{\Phi} \in \mathbb{R}^{N \times N_b}$ and $a \in \mathbb{R}^{N_b}$ with $K$ non-zero elements. In the sparse reconstruction formulation above, $\boldsymbol{\Phi} \in \mathbb{R}^{N \times N_b}$ is used instead of $\boldsymbol{\Phi} \in \mathbb{R}^{N \times K}$ as the sparsity of the system $K$ is not known *a priori*. Consequently, a more exhaustive basis set of dimension $N_b \approx P > K$ is typically employed. To represent $N$-dimensional data, one can atmost use $N$ basis vectors, i.e., $N_b \leq N$. In practice, the number of possible basis need not be $N$ and can be represented by $N_b \ll N$ as only $K$ of them are needed to represent the acquired signal up to a desired quality. This is typically the case when $\boldsymbol{\Phi}$ is composed of optimal data-driven basis vectors such as POD modes. The reconstruction problem is then recast as identification of these $K$ coefficients. In many practical situations, the knowledge of $\Phi$ and $K$ is not known *a priori* and $N_b, N$ are typically user input. Standard *transform coding* [25] practice in image compression involves collecting a high resolution sample, transforming it to a Fourier or wavelet basis space where the data is sparse and retain the $K$-sparse structure while discarding the rest of the information. This is the basis of JPEG and JPEG-2000 compression standards [25]. The *sample and then compress* mechanism still requires acquisition of high resolution samples and processing them before reducing the dimensionality. This is highly challenging as handling large amounts of data is difficult in practice due to demands on processing power, storage, and time. Compressive sensing [14, 16, 38, 39, 40] focuses on direct sparse sensing based inference of the $K$-sparse coefficients by essentially combining the steps in equations 1 and 3 as below:

$$\tilde{X} = C\boldsymbol{\Phi} a = \boldsymbol{\Theta} a, \tag{4}$$

where $\boldsymbol{\Theta} \in \mathbb{R}^{P \times N_b}$ is the map between the basis coefficients $a$ that represent the data in a feature space and the sparse measurements, $\tilde{X}$ in physical space. The challenge in solving for $X$ using the underdetermined equation 1 is that $C$ is ill-conditioned and $X$ in itself is not sparse. However, when $X$ is sparse in $\boldsymbol{\Phi}$, the reconstruction using $\boldsymbol{\Theta}$ in equation 4 becomes practically feasible by solving for $a$ that is $K$-sparse. Thus, one effectively solves for $K$ unknowns using $P$ constraints and this is typically achieved by computing a sparse solution $a$ as per equation 7 by minimizing the corresponding $s$-norm. $X$ is then recovered from equation 3. $s$ chosen as 2 represents the $l_2$ norm reconstruction of $X$ and gives the $a$ with least energy. The $l_2$-based method can be solved by a minimization problem as shown in Equation 5.

$$\min \left\lVert \tilde{X} - \boldsymbol{\Theta} a \right\rVert_2^2.$$ (5)

Using left pesudo-inverse of $\boldsymbol{\Theta}$, Equation 5 becomes:

$$a = (\boldsymbol{\Theta})^+ \tilde{X},$$ (6)

where $\boldsymbol{\Theta}^+$ can be approximated as a solution to the normal equation as $\left(\boldsymbol{\Theta}^T \boldsymbol{\Theta}\right)^{-1} \boldsymbol{\Theta}^T \tilde{X}$. This least-squares solution procedure is nearly identical to the original GPOD algorithm developed by Everson and Sirovich [23] if $\boldsymbol{\Phi}$ is chosen as the POD basis. However, $\tilde{X}$ in GPOD contains zeros as placeholders for all the missing elements whereas the above formulation retains only the measured data points. The GPOD formulation summarized in section 2.4 is plagued by issues that are beyond the scope of this article. Unfortunately, this $l_2$ approach rarely, if ever finds the $K$-sparse solution. A natural way to enhance sparsity of $a$ is to minimize $\lVert a' \rVert_0$, i.e., minimize the number of non-zero elements such that $\boldsymbol{\Theta} a' = \tilde{X}$ is satisfied. It has been shown [41] that with $P = K + 1$ ($P > K$ in general) *independent* measurements, one can recover the sparse coefficients with high probability using $l_0$ reconstruction. This condition can be heuristically interpreted as each measurement needing to excite a different basis vector $\phi_i$ so that its coefficient $a_i$ can be optimally identified. If two or more measurements excite the same basis $\phi_j$ then additional measurements may be needed to produce acceptable reconstruction. On the other hand, for $P \leq K$ independent measurements, the probability of recovering the sparse solution is highly diminished. Nevertheless, $l_0$-minimization is a computationally complex, $np$-hard and poorly conditioned problem with no stability guarantees.

$$a = \operatorname{argmin} \quad \lVert a' \rVert_{0_l} \quad \text{such that} \quad \boldsymbol{\Theta} a' = \tilde{X}.$$ (7)

The popularity of compressed sensing arises due to the theoretical advances [42, 43, 44, 45] guaranteeing near-exact reconstruction of the uncompressed information by solving for the $K$ sparsest coefficients. The $l_1$ reconstruction is a relatively simpler convex optimization problem and solvable using linear programming techniques for basis pursuit [46, 14, 38]. Theoretically, one can perform the traditional brutal search to locate the largest $K$ coefficients of $a$, but the computational effort increases exponentially with the dimension. To overcome this burden, a host of greedy algorithms [15, 17, 47] have been developed to solve the $l_1$ minimization problem in Equation 7 with complexity $\mathcal{O}(N^3)$ for $N_b \approx N$. However, the price one pays here is that $P > \mathcal{O}(K log(N/K))$ measurements are needed [14, 38, 42] to *exactly* reconstruct the K-sparse vectors using this approach. The schematics of both $l_2$ and $l_1$-based formulations are illustrated in Figure 1.

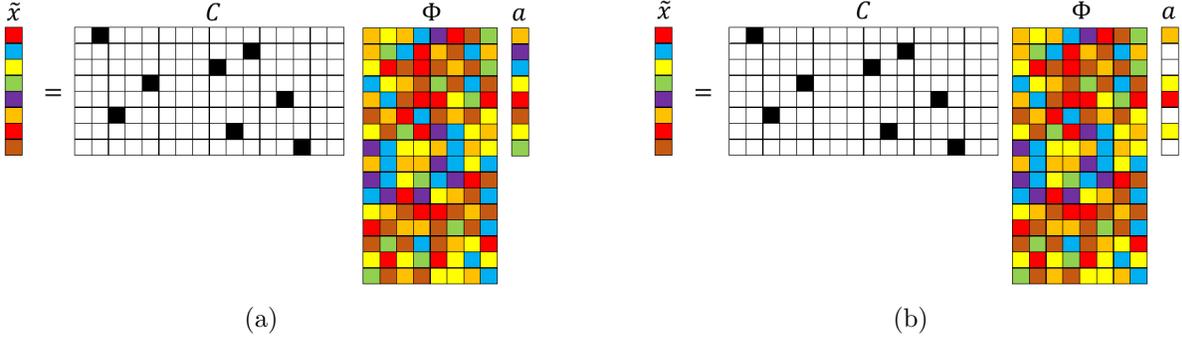

Figure 1: Schematic illustration of $l_2$ (left) and $l_1$ (right) minimization reconstruction for sparse recovery using a single-pixel measurement matrix. The numerical values in $C$ are represented by colors: black (1), white (0). The other colors represent numbers that are neither 0 nor 1. In the above schematics $\tilde{X} \in \mathbb{R}^P$, $C \in \mathbb{R}^{P \times N}$, $\mathbf{\Phi} \in \mathbb{R}^{N \times N_b}$ and $a \in \mathbb{R}^{N_b}$, where $N_b \leq N$. The number of colored cells in $a$ represents the system sparsity $K$. $K = N_b$ for $l_2$ and $K < N_b$ for $l_1$.

In summary, there are three parameters $N_b, K, P$ that impact the reconstruction framework. $N_b$ represents the candidate basis space dimension employed for this reconstruction and can at worst obey $N_b \approx N$. $K$ represents the desired system sparsity and tied to the desired quality of reconstruction. That is, $K$ is chosen such that if these features are predicted accurately, then the achieved reconstruction is satisfactory. The more sparse a system, the smaller $K$ is for a desired reconstruction quality. $P$ represents the available quantity of sensors provided as input to the problem. The interplay of $N_b, K$, and $P$ determines the choice of the algorithm employed, i.e., whether the reconstruction is based on $l_1$ or $l_2$ minimization, and the reconstruction quality as summarized in Table 1. In general, $K$ is not known *a priori* and is tied to the desired quality of reconstruction. $N$ and $N_b$ are chosen by the practitioner and depends on the feature space in which the reconstruction will happen. $N$ is the preferred dimension of the reconstructed state, and $N_b$ is the dimension of the candidate basis space in where the reconstruction problem is formulated. As shown in Table 1, for the case with $K = N_b$, the best reconstruction will predict the $K$ weights correctly (using $l_2$ for the over determined problem) and can be as bad as $P$ when $P < K$ (using $l_1$ minimization for the under determined problem). In all the cases explored in this discussion, the underlying assumption of $N_b \ll N$ is used. When $K < N_b$ and $N_b > P \geq K$, the worst case prediction will be $K$ weights (for a desired sparsity $K$) as compared to $P$ weights for the best case (maximum possible sparsity) using $l_1$ minimization. With $K < N_b$ and $N_b > K > P$, the best case reconstruction will be $P$ weights using $l_1$. For $P \geq N_b > K$, the best reconstruction will predict $N_b$ weights as compared to $K$ for the worst case. Thus in cases $1, 4, 5$ the desired reconstruction sparsity is always realized whereas in cases 2 and 3, the sensor quantity determines the outcome.

All of the above sparse recovery estimations are conditional upon the measurement basis (rows of $C$) being incoherent with respect to the sparse basis $\mathbf{\Phi}$. In other words, the measurement basis cannot sparsely represent the elements of the "data basis." This is usually accomplished by using a random sampling for sensor placement, especially when $\mathbf{\Phi}$ is made up of Fourier functions or wavelets. If the basis functions $\Phi$ are orthonormal, such as wavelet and POD basis, one can discard the majority of the small coefficients in $a$ (setting them as zeros) and still retain reasonably accurate reconstruction. The mathematical explanation of this conclusion has been previously shown in [16]. However, it should be noted that incoherency is a necessary, but not sufficient condition for exact reconstruction. Exact reconstruction requires optimal sensor placement to capture the most information for a given flow field. In other words, incoherency alone does not guarantee optimal reconstruction which also depends on the sensor placement as well as quantity.

Table 1: The choice of sparse reconstruction algorithm based on problem design using parameters $P$ (sensor sparsity), $K$ (targeted system sparsity) and $N_b$ (candidate basis dimension).

| Case | $K - N_b$ Relationship | $P - K$ Relationship | Algorithm | Reconstructed Dimension |
|------|------------------------|----------------------|-----------|-------------------------|
| 1 | $K = N_b$ | $P \geq K$ | $l_2$ | $K$ |
| 2 | $K = N_b$ | $P < K$ | $l_1$ | $P$ |
| 3 | $K < N_b$ | $P < K < N_b$ | $l_1$ | $P$ |
| 4 | $K < N_b$ | $N_b > P \geq K$ | $l_1$ | $K$ or $P$ |
| 5 | $K < N_b$ | $P \geq N_b > K$ | $l_2$ | $K$ or $N_b$ |

## 2.2. Data-driven Sparse Basis Computation using POD

In the SCR framework, basis such as POD modes, Fourier functions, and wavelets [14, 16] can be used to generate low-dimensional representations for both $l_2$ and $l_1$-based methods. While an exhaustive study on the effect of different choices on reconstruction performance is potentially useful, in this study we refer to [6] for a flavor of such assessment. It is shown that data-driven POD basis outperform generic cosine basis when reconstructing with small amounts of data while the accuracy becomes comparable with more data. As a result, we only focus on the use of POD modes as the basis in this reconstruction study, i.e., the basis vectors constituting $\mathbf{\Phi}$ are POD modes in the rest of this paper.

Proper orthogonal decomposition (POD), also known as Principal Components Analysis (PCA) or Singular Value Decomposition (SVD), is a dimensionality reduction technique that computes a linear combination of low-dimensional basis functions (POD modes) and weights (POD coefficients) from snapshots of experimental or numerical data [2, 4] through eigendecomposition of the spatial correlation tensor of the data. It was introduced in the turbulence community by Lumley [48] to extract coherent structures in turbulent flows. The resulting singular vectors or POD modes represent an orthogonal basis that maximizes the energy capture from the flow field. For this reason, such eigenfunctions are considered *optimal* in terms of energy capture and other optimality constraints are theoretically possible. Taking advantage of the orthogonality, one can project these POD basis onto each snapshot of data in a Galerkin sense to deduce coefficients that represent evolution over time in the POD feature space. The optimality of the POD basis also allows one to effectively reconstruct the full field information with knowledge of very few coefficients, a feature that is attractive for solving sparse reconstruction problems such as in Equation 4. Since the eigendecomposition of the spatial correlation tensor of the flow field requires handling a system of dimension $N$, it requires significant computational expense. An alternative method is to compute the POD modes using the method of snapshots [49] where the eigendecomposition problem is reformulated in a reduced dimension (assuming the number of snapshots in time is smaller than the spatial dimension) framework as summarized below. Consider that $X \in \mathbb{R}^{N \times M}$ is the full field representation with only the fluctuating part, i.e., the temporal mean is taken out of the data. $N$ is the dimension of the full field representation and $M$ is the number of snapshots. The procedure involves computation of the temporal correlation matrix $\bar{C}_M$ as:

$$\bar{C}_M = X^T X. \tag{8}$$

The resulting correlation matrix $\bar{C}_M \in \mathbb{R}^{M \times M}$ is symmetric and an eigendecomposition problem can be formulated as:

$$\bar{C}_M V = V \Lambda , \tag{9}$$

where the eigenvectors are given in $V = [v_1, v_2, ..., v_M]$ and the diagonal elements of $\Lambda$ are the eigenvalues $[\lambda_1, \lambda_2, ..., \lambda_M]$. Typically, both the eigenvalues and corresponding eigenvectors are sorted in descending order such as $\lambda_1 > \lambda_2 > ... > \lambda_M$. The POD modes $\Phi$ and coefficients $a$ can then be computed as

$$\Phi = XV\Lambda^{-1}. \tag{10}$$

One can represent the field $X$ as a linear combination of the POD modes $\Phi$ as shown in equation 3 and leverage orthogonality, i.e., $\Phi^{-1} = \Phi^T$, to directly estimate the POD as shown in Equation 11,

$$a = \Phi^T X. \tag{11}$$

It is worth mentioning that subtracting the temporal mean from the input data is not always necessary for the above procedure. Further, the snapshot procedure fixes the maximum number of POD basis vectors to $M$ which is typically much smaller than the dimension of the full state vector, $N$. If one wants to reduce the dimension further, then a criterion based on energy capture can be devised so that the modes carrying the least amount of energy can be *truncated*. For many common fluid flows, using the first few POD modes and coefficients are sufficient to capture almost all the relevant dynamics. However, for turbulent flows with large-scale separation, a significant number of POD modes will need to be retained.

*2.3. Measurement Locations, Data Basis and Incoherence*

We recall from subsection 2.1 that the reconstruction performance estimates are contingent upon the measurement matrix, $C$ being *necessarily* incoherent with respect to the sparse basis $\mathbf{\Phi}$ [16] and this is usually accomplished by employing a random sampling for the sensor placement. In practice, one can adopt two types of sparse representation of the data, namely, single-pixel measurement [7, 50] or random projections [39, 14, 16]. Typically, single-pixel measurement refers to measuring information at the particular spatial locations. One of the most practical applications using single-pixel is through mobile unmanned systems (UAS) sensing of atmospheric boundary layer flows or weather turbulence. Another popular choice of sensing method in the compressive sensing or image processing community is random projections where the compression matrix is populated using normally distributed random numbers on to which the full state data is projected. As per theory, the random matrix is highly likely to be incoherent with any fixed basis [16], and hence efficient for sparse recovery purposes. However, for most of the fluid flow applications, the sparse data is usually sourced from point measurements and hence, the single-pixel approach is practically relevant.

Irrespective of the approach adopted, the measurement matrix $C$ and basis functions $\mathbf{\Phi}$ should be incoherent to ensure optimal sparse reconstruction. This essentially implies that one should have sufficient measurements distributed in space to excite the different modes relevant to the data being reconstructed. Equation 12 can be used to estimate the extent of coherency between $C$ and $\mathbf{\Phi}$ in the form of an *coherency* number, $\mu$ [51],

$$\mu(C, \mathbf{\Phi}) = \sqrt{N} \cdot \max_{i \leq P, \ j \leq K} |\langle c_i, \phi_j \rangle| \,, \tag{12}$$

where $c_i$ is a row vector in $C$ and $\phi_j$ is a column vector of $\mathbf{\Phi}$. $\mu$ typically ranges from 1 (incoherent) to $\sqrt{N}$ (coherent). The smaller the $\mu$, the less measurements one needs to reconstruct the data in an $l_1$ sense. This is because the coherency parameter enters as the prefactor in the minimum sensor estimate in $l_1$-based CS.

To simplify the sensor placement strategy, the *Matlab* function randperm(N) is used to generate random permutation from 1 to $N$ and the first $P$ values are chosen as the smapling locations in the data. In practice,the sensor locations are available as input. Other sensor placement algorithms such as K-means clustering, the data-driven Online sparse Gaussian Processes [52] and physics-based approaches [53] are also available. Since the experiments performed in this study have a relatively low number of sparse measurements as compared to the total number of grid points, i.e., $P \ll N$, we will only focus on randomly single-pixel measurements. It should be noted that although the measurements are incoherent, the sensor placement need not be optimal for reconstruction. Optimal sensor placement methods are numerous and can be based on physics [53] or reconstruction performance for a given convolution basis (kernel) [52]. Review methods for optimal sensor placement. A thorough study on the role of sensor placement on reconstruction quality is much needed and an active topic of research, but not considered within the scope of this work. However, a preliminary assessment by changing the seed of the random number generator for identifying sensor locations and their impact on reconstruction quality is carried out and results summarized in section 4.

To examine the coherency of our sensor placement design as seen from the measurement matrix with basis functions $\phi$, $\mu$ is computed for select cases at different $Re$ numbers as summarized in Table 2. Given that the fluid flows considered in this article are multi-dimensional, a coherency number, $\mu$, for each flow velocity component is computed. We observe that for all the cases considered here, $\mu_u$ and $\mu_v$ are relatively small. In this study, we consider sparse reconstruction of cylinder wake flows at multiple $Re$ numbers, $Re = 100, 300, 800, 1000$. Considering that $\sqrt{N} = 154.713$ for $Re = 100$ and $\sqrt{N} = 308.181$ for $Re = 300, 800, 1000$, our $\mu$ values are $O(1)$ and hence, indicating low coherency between $C$ and $\Phi$. In essence, the single-pixel approach with random sensor placement used in this analysis has proven to be incoherent which. If low coherency is met, $C\Phi$ also expected to satisft the *restricted isometry principle* (RIP) [54] as shown below:

$$(1 - \delta) \, ||a||_2^2 \leq ||C\Phi a||_2^2 \ \leq (1 + \delta) \, ||a||_2^2 \,, \tag{13}$$

where $\delta$ is a restricted isometry constant, and $a$ is any $K$-sparse vector with the same non-zero values as the reconstructed $K$-sparse solution. In practice, the RIP is impractical for use as a verification tool and the coherency number estimation offers a way out. How coherency of the measurements with the underlying basis vectors impact sparse reconstruction accuracy is examined in the later sections.

Table 2: The coherency number $\mu$ for each $Re$ number. Subscripts $u$ and $v$ correspond to the horizontal and vertical velocity components, respectively.

| $Re$ | 100 | 300 | 300 | 800 | 800 | 800 | 1000 | 1000 | 1000 | 1000 |
|---|---|---|---|---|---|---|---|---|---|---|
| Seed | 101 | 101 | 200 | 101 | 101 | 132 | 101 | 101 | 101 | 132 |
| $P$ | 20 | 20 | 12 | 30 | 24 | 24 | 40 | 40 | 24 | 24 |
| $K$ | 7 | 10 | 10 | 15 | 15 | 15 | 4 | 20 | 15 | 15 |
| $\mu_u$ | 1.471 | 3.050 | 2.6872 | 4.372 | 4.372 | 6.264 | 2.285 | 5.467 | 3.8829 | 6.001 |
| $\mu_v$ | 1.136 | 6.050 | 4.0762 | 6.203 | 4.968 | 5.426 | 3.578 | 6.592 | 4.8341 | 5.444 |

### 2.4. The GPOD Algorithm as an Alternate Formulation for Sparse Convolution-based Reconstruction

The Gappy POD or GPOD algorithm of [23] can be viewed as an $l_2$ minimization reconstruction of the sparse recovery problem summarized through equations 4,5 and 6 with $\mathbf{\Phi}$ composed of $K \leq M$ POD basis vectors, i.e. dimension of $a$ is $K \leq M$. In the GPOD one normally chooses $K$ and by default it is $M$. However, the primary difference between the SCR framework in equations 4 and GPOD as shown in equation 15 is the construction of the measurement matrix $C$ and the sparse measurement vector $\tilde{X}$. In SCR (equation 4) $\tilde{X} \in \mathbb{R}^P$ is a *compressed* version containing only the measured data, while in the GPOD framework, $\tilde{X} \in \mathbb{R}^N$ is a *masked* version of the full state vector. Therefore, $\tilde{X}$ in GPOD is composed of $P$ measured values with the rest being zeroed out or more generally, a filtered version of $X$. Given a complete set of data $X \in \mathbb{R}^N$, its POD basis functions $\phi_k \in \mathbb{R}^N$ and associated coefficients $a \in \mathbb{R}^K$ can be expressed as:

$$X = \sum_{k=1}^{K} a_k \phi_k. \tag{14}$$

The masked (incomplete) data vector $\tilde{X} \in \mathbb{R}^N$, measurement matrix $C$ and mask vector $m \in \mathbb{R}^N$ are related by:

$$\tilde{X} = <m \cdot X> = CX, \tag{15}$$

where $C \in \mathbb{R}^{N \times N}$. Contrastingly in SCR, $C \in \mathbb{R}^{P \times N}$, whereas in GPOD, $C \in \mathbb{R}^{N \times N}$ resulting in a larger matrix with numerous rows of zeros as shown in fig. 2. The subtle differences between the formulation of the sparse recovery problem in GPOD and SCR can be realized by comparing the algorithm in Figure 2 with the $l_2$ and $l_1$-based SCR methods shown in Figure 1. In order to bypass the complexity of handling the $N \times N$ matrix, a mask vector, $m$ of size $N \times 1$ with elements in the form of 1s and 0s operates on $X$ through a point-wise multiplication operator $< \cdot >$. As an illustration, the point-wise multiplication is represented as $\tilde{X}_i = <m_i \cdot X_i>$ for each snapshot $i = 1..M$ where each element of $X_i$ multiplies with the corresponding element of $m_i$. Each data snapshot, $X_i$ can have its own measurement masj $m_i$ which is a useful way to represent the evolution of sparse sensor locations over time. The SCR formulation (eq. 4) can also support time varying sensor placement, but would require a compression matrix, $C_i$ that is unique for each snapshot. This approach is by design much more computationally and storage intensive. The goal of the GPOD (and SCR) procedure is to recover the full data by approximating the POD coefficients $\bar{a}$ (in the $l_2$ sense) with POD modes, $\Phi_K$, learnt offline from training data (usually these are snapshots of the full field data):

$$\tilde{X} \approx m \sum_{k=1}^{K} \bar{a}_k \phi_k. \tag{16}$$

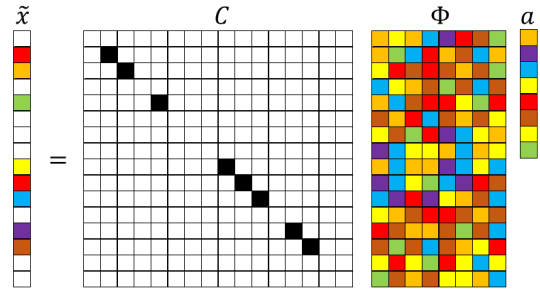

Figure 2: Schematic of GPOD or POD-SCR algorithm for sparse reconstruction problem. The corresponding numerical values for colored blocks: black (1), white (0), others (arbitrary number).

Unlike the standard POD framework, the coefficient vector $\bar{a}$ cannot be computed directly by projection onto the basis vector using the inner product, i.e., using equation 14 and the projection $a = \langle \mathbf{\Phi}, X \rangle$ as the POD modes, $\mathbf{\Phi}$ are not designed to optimally represent the incomplete data $\tilde{X}$. The solution procedure is to obtain the "best" approximation of the coefficient $\bar{a}$, by minimizing the error $\mathcal{E}$ in the $l_2$ sense:

$$\mathcal{E} = \left\| \tilde{X} - m \sum_{k=1}^{K} \bar{a}_k \phi_k \right\|_2^2 = \left\| \tilde{X} - m \cdot \mathbf{\Phi}\bar{a} \right\|_2^2 = \left\| \tilde{X} - C\mathbf{\Phi}\bar{a} \right\|_2^2. \tag{17}$$

The above equation clearly denotes the connection between GPOD and SCR when the mask vector is non-compactly represented by $C \in \mathcal{R}^{N \times N}$. In equation 17 we see that $m$ acts on each column of $\mathbf{\Phi}$ through a point-wise multiplication operator which is equivalent to masking each basis vector $\phi_k$. We note that the above formulation is valid only for a single snapshot reconstruction. In the case of multiple snapshots, each denoted by $i = 1..M$, the mask vector is $m_i$, the masked state vector is $\tilde{X}_i$ and the error $\mathcal{E}_i$ represents the single snapshot reconstruction error that will be minimized to compute the approximate POD features $\bar{a}^i$. It can easily be seen from below that one can minimize the different $\mathcal{E}_i$'s simultaneously to learn the POD feature matrix, $\bar{\boldsymbol{a}} \in \mathcal{R}^{K \times M}$. Denoting the masked basis functions as $\tilde{\phi}_k(z) = < m(z) \cdot \phi_k(z) >$, eq. 17 can be re-written as:

$$\mathcal{E} = \left\| \tilde{X} - \sum_{k=1}^{M} \bar{a}_k \tilde{\phi}_k \right\|_2^2. \tag{18}$$

In the above formulation, $\tilde{\mathbf{\Phi}}$ is analogous to $C\mathbf{\Phi} = \mathbf{\Theta}$ in eq. 4. To minimize $\mathcal{E}$, one computes the derivative with respect to $\bar{a}$ and equated to zero as below:

$$\frac{\partial}{\partial \bar{a}}(\mathcal{E}) = 0. \tag{19}$$

The result is a linear system or commonly known as the normal equations:

$$\mathbf{M}\bar{a} = f, \tag{20}$$

where $\mathbf{M}_{k1,k2} = \langle \tilde{\phi}_{k1}, \tilde{\phi}_{k2} \rangle$ or $\mathbf{M} = \tilde{\mathbf{\Phi}}^T \tilde{\mathbf{\Phi}}$ and $f_k = \langle \tilde{X}, \tilde{\phi}_k \rangle$ or $f = \tilde{\mathbf{\Phi}}^T \tilde{X}$. The resulting reconstructed solution is:

$$\bar{X} = \sum_{k=1}^{M} \bar{a}_k \phi_k. \tag{21}$$

Algorithm 1 summarizes the above GPOD-type SR algorithm assuming the POD basis functions ($\phi_k$) are known. The above solution procedure of sparse recovery is the same as that described in section 2.1 except for the dimensions of the $\tilde{X}$ and $C$. In addition, there exists another difference at the final step: the final form of the repaired data requires one to overwrite the reconstructed data with known sparse data at the measurement locations. However, this is expected to impact the overall reconstruction minimally as it zeros out the error only at the sparse sensor locations. If the basis functions ($\phi_k$) are unavailable, then algorithm 1 can be extended to an iterative procedure [18] that simultaneously solves for $\mathbf{\Phi}$ and $\bar{a}$. In such methods one can compute an intial guess for the basis functions by using an ensemble average based initial solution field [18] or by Kriging ideas [21]. In this study, we use the above algorithm 1 on account of its computational efficiency to perform the $l_2$-based sparse reconstruction of cylinder wakes at different flow regimes characterized by $Re$ number.

---

**Algorithm 1:** $l_2$-based algorithm: GPOD sparse reconstruction with known basis, $\boldsymbol{\Phi}$.

> **input** : Full data ensemble $X \in \mathbb{R}^{N \times M}$
>   Incomplete data vector $\tilde{X} \in \mathbb{R}^N$
>   the mask vector $m \in \mathbb{R}^N$.
>
> **output:** Approximated full data vector $\bar{X} \in \mathbb{R}^N$

**1** Option: take out the temporal mean of the ensemble $X$.
**2** Compute SVD of $X$ to obtain the POD basis function $\boldsymbol{\Phi}$.
**3** Decide on number of modes to retain.
**4** Build a least square problem: $\mathbf{M}\bar{a} = f$.
**5** Compute masked basis function: $\tilde{\boldsymbol{\Phi}} = m\boldsymbol{\Phi}$.
**6** Compute matrix $\mathbf{M} = \tilde{\boldsymbol{\Phi}}^T \cdot \tilde{\boldsymbol{\Phi}}$.
**7** Compute vector: $f = \tilde{\boldsymbol{\Phi}}^T \cdot \tilde{X}$.
**8** Solve $\bar{a}$ from the least squares problem: $\mathbf{M}\bar{a} = f$.
**9** Reconstruct the approximated solution $\bar{X} = \boldsymbol{\Phi}\bar{a}$.
**10** Substitute the gappy data back to $\bar{X}$:
    (a) $\bar{X}_i = \bar{X}_i$   if   $m_i = 0$
    (b) $\bar{X}_i = \tilde{X}_i$   if   $m_i = 1$
**11** Output the approximated full data vector $\bar{X}$.

---

## 3. Data Generation for Cylinder Wake Flows

Studies of cylinder wakes [55, 56, 57, 31] have attracted considerable interest from the model reduction community for its particularly rich flow physics content, encompassing many of the complexities of nonlinear dynamical systems, while easy to simulate accurately on the computer using established CFD tools. In this study, we explore data-driven sparse reconstruction for he cylinder wake flow at multiple $Re$ numbers, $Re = 100, 300, 800,$ and $1000$ and at two separate temporal regions of interest: the periodic phase (limit cycle) and the transient phase (evolution towards limit cycle). To generate two-dimensional cylinder flow data, we adopt the spectral Galerkin method [58] to solve incompressible Naiver-Stokes equations, as shown in Eq. 22$a$ below:

$$\frac{\partial u}{\partial x} + \frac{\partial u}{\partial y} = 0, \tag{22a}$$

$$\frac{\partial u}{\partial t} + u\frac{\partial u}{\partial x} + v\frac{\partial u}{\partial y} = -\frac{\partial P}{\partial x} + \nu\nabla^2 u, \tag{22b}$$

$$\frac{\partial v}{\partial t} + u\frac{\partial v}{\partial x} + v\frac{\partial v}{\partial y} = -\frac{\partial P}{\partial y} + \nu\nabla^2 v, \tag{22c}$$

where $u$ and $v$ are horizontal and vertical velocity components. $P$ is the pressure field, and $\nu$ is the fluid viscosity. The rectangular domain used for this flow problem is $-25D < x < 45D$ and $-20D < y < 20D$, where $D$ is the diameter of the cylinder. For the purposes of this study, data from a reduced domain, i.e., $-2D < x < 10D$ and $-3D < y < 3D$, is used. The mesh was designed to sufficiently resolve the thin shear layers near the surface of the cylinder and transit wake physics downstream. For the case of $Re = 100$ the grid includes $24,000$ points whereas for $Re = 300, 800,$ and $1000$ the grid is refined to include approximately $95,000$ points for the sampled flow region. The computational method employed is a fourth order spectral expansion within each element in each direction. The sampling rate for each snapshot output is chosen as $\Delta t = 0.2$ seconds.

## 4. Sparse Reconstruction of Canonical Fluid Flows

### 4.1. Cylinder Wake Limit-cycle Dynamics

In this section, we explore sparse reconstruction of fluid flows at multiple $Re$ numbers using the GPOD algorithm for the cylinder flow with well-developed periodic vortex shedding behavior. For examples, time-series snapshots of stream-wise velocity component ($u$) in color contour are shown in Figure 3 for $Re = 100, 300, 800$ and $1000$. It is seen that the length-scale of the vortex shredding decreases with $Re$ number, $Re$ and therefore, requiring finer spatial resolutions.

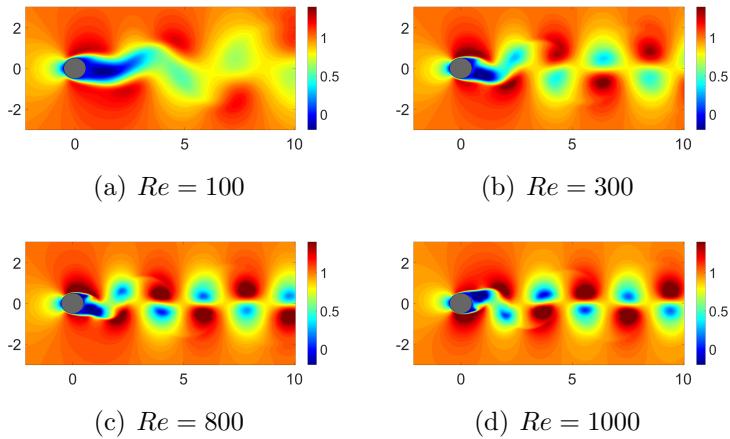

(a) $Re = 100$      (b) $Re = 300$

(c) $Re = 800$      (d) $Re = 1000$

Figure 3: Color contour snapshots of the stream-wise velocity component for the cylinder flow at different $Re$ number at arbitrary time unit $T$.

In this study, we choose 300 snapshots of data for each $Re$, corresponding to a non-dimensional time ($T = \frac{Ut}{D}$) of $T = 60$ with uniform temporal spacing of $dT = 0.2s$. $T = 60$ corresponds to multiple ($\approx 10 - 15$) cycles of periodic vortex shedding behavior for the flows with $Re = 100 - 1000$. The first three POD modes and coefficients are computed and shown in Figure 15 (in the Appendix) and 4, respectively. Similar to the velocity field visualized in figure 3, the dominant POD modes (mode 1 and mode 2) capture the symmetric vortex shedding patterns at various length scales for all the cases. The vestiges of the onset of instability at the higher $Re$ numbers ($Re = 300$, 800, and 1000) is observed from the asymmetry in mode 3. This is consistent with the observations in [59], where the laminar vortex shedding happens at around $Re = 47$, and becomes unstable at $\sim 190$ which deforms the small-scale structures at higher POD modes. On the other hand, the temporal evolution of POD coefficients show periodic limit-cycle behavior for all the $Re$ numbers with the characteristic frequencies increasing with $Re$. The dependence of this $Re$ dependent dynamics on sparse reconstruction performance is explored below.

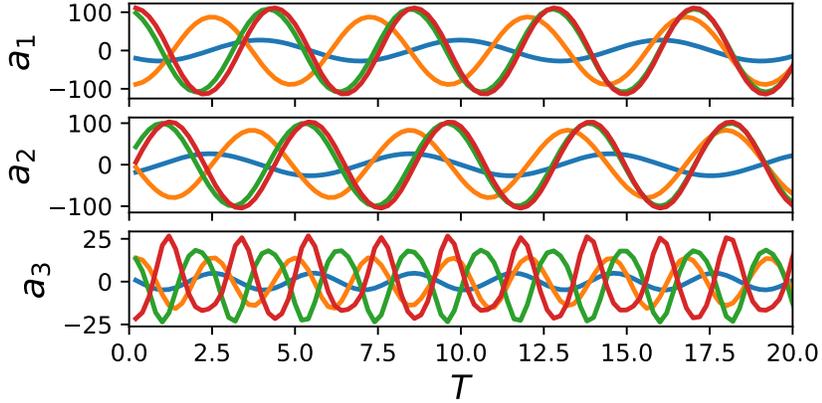

Figure 4: Time evolution of the first three POD coefficients (Top: $a_1$; Middle: $a_2$; Bottom: $a_3$) for the cylinder flow at different $Re$ number ($Re = 100$ (Blue), 300 (Orange), 800 (Green), 1000 (Red)).

## 4.2. Sparse Reconstruction Experiments and Analysis

To assess the SCR performance we study the reconstruction of simulation data where the full field representation is known *a priori*. The sparse sensor locations are chosen as single point measurements using a random sampling of the full field data. The reconstruction performance is evaluated by comparing the field predicted from CFD simulation with that from SCR. This approach is chosen so as to assess the relative roles of the choice of system sparsity ($K$), sensor sparsity ($P$) and sensor placement ($C$). In particular, we aim to accomplish the following through this study: (i) check if $P > K$ is a necessary condition for accurate POD-based reconstruction of the fluid flow across $Re$ numbers and flow regimes; (ii) estimation of sparsity metric, $K$ for desired reconstruction quality for the flows considered in this study; (iii) assess role of incoherence of sensor placement on reconstruction and (iv) establish whether the above conclusions for limit-cycle data are relevant to transient wake flow dynamics.

The first step of the POD-based sparse reconstruction framework is to identify the basis from the data using the method of snapshots [49] and as shown in equations 8-11. For the sparse reconstruction of cylinder flow described in section 4.1, the POD basis, eigenvalues and coefficients are obtained from the full data ensemble, i.e., $M = 300$ snapshots corresponding to $T = 60$ non-dimensional times. This gives rise to $M$ POD basis for use in the reconstruction process in equation 3, i.e. a candidate basis dimension of $N_b = M$. As shown in Table.1, the choice of algorithms depends on the combination of system sparsity ($K$), data sparsity ($P$) and dimension of the candidate basis space, $N_b$. Recalling from the earlier discussion in section 2, we see that $P \geq K$ would always require an $l_2$ method for a desired reconstruction sparsity $K$ as long as $P \geq N_b$. Data-driven basis such as POD are energy optimal for the training data and hence, contain *built-in sparsity*. That is, as long as the basis is relevant for the flow to be reconstructed, retaining only the most energetic modes (basis) should generate the best possible reconstruction for the given sensor locations. This is in contrast to most reconstruction problems where the hierarchy of energy-relevant basis to the sparse data is not known *a priori* and therefore require a larger candidate basis space to search from, i.e. $N_b > P$. In the unique case of data-driven POD basis, the basis vectors are ordered on decreasing energy content and therefore, retaining the first $K$ desired modes will also be the $K-$sparse solution from CS ($l_1$) as long as the basis is relevant. In other words, a way to identify the sparse basis set of dimension $K$ for a relevant sparse-data is automatically encoded in the POD framework. In the following subsection 4.3, we verify this by comparing $l_2$ reconstruction for $K = N_b < M$ with $l_1$ reconstruction for $K < N_b = M$. If true, this allows one to bypass the need for more computationally expensive $l_1$ methods in such special cases. However, if one were to build a library of POD basis across different flow regimes or reconstruct from sparse data that may not be similar to the training data used to generate the POD basis, then the candidate basis needs to encompass a larger subspace with $N_b >> K$ so that one could search for an optimally sparse ($l_1$) solution.

### 4.3. Comaprison of Energy and $l_1$ Sparse Reconstruction

In this subsection, we verify the equivalence between energy- and $l_1-$sparse reconstruction for the cylinder wake data considered in this study. The success of this verification will imply that the chosen SR POD basis based on energy content in the training data also efficiently spans the data being reconstructed. Energy sparsity is trivial to implement as long as one knows the pecking order of each mode as provided by the data-driven POD-basis and the data to be reconstructed is spanned by this basis space. Consequently, one can predetermine the $K$-sparse basis or most energetic $K$ modes, set $K = N_b$ and perform a $l_2$ reconstruction on this basis space provided sufficient sparse data, $P > K$, is available. To verify this, we consider two cases of reconstruction for the limit-cycle cylinder wake at $Re = 100$. Case (a): $P > K$,

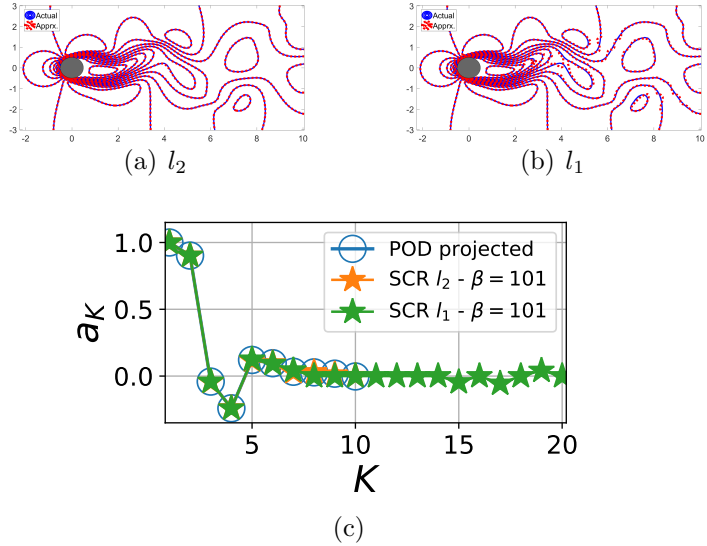

(a) $l_2$      (b) $l_1$

(c)

Figure 5: Comparison of the sparse reconstruction using both $l_1$ and $l_2$ minimization methods for basis that is ordered in terms of energy content. The reconstructed and actual flowfields at $T = 0.2$ are compared in (a) for $l_2$ and (b) $l_1$. The corresponding POD features from both methods are shown in (c).

$K = N_b$ and employs $l_2$ reconstruction; Case (b): $P > K$, $K < N_b \leq M$ and employs $l_1$ reconstruction. Figure 5 and table 3 compares the reconstructed fields and the $K$ predicted weights for the two cases for a single snapshot corresponding to $T = 0.2$. In this study we use a greedy optimal matching pursuit algorithm, CoSAMP as described by Needell and Tropp in [17].

Table 3: caption

|       | $Re$ | $K$ | $P$ | $N_b$ | $\beta$ |
|-------|------|-----|-----|-------|---------|
| $l_2$ | 100  | 10  | 40  | 10    | 101     |
| $l_1$ | 100  | 10  | 40  | 20    | 101     |

While there exists a small amount of error in the $l_1$ predictions, the results confirms equivalence of these two approaches for POD-based (energy sparse) reconstruction of cylinder wake flows considered in this study. It is worth noting that $l_1$ should be the preferred choice when the relevance of sparse data to the candidate basis space is in doubt. For the rest of this paper, we will leverage energy sparsity of the data with respect to the candidate POD basis and adopt the problem formulation with $K = N_b$ along with an $l_2$ minimization algorithm.

### 4.4. Sparsity and Energy Metrics

For this energy-based SR we explore the conditions for accurate recovery of information in terms of data sparsity $(P)$ and system sparsity $(K)$. As long as the measurements are incoherent with respect to the basis $\mathbf{\Phi}$ and the system is overdetermined, i.e., $P > K$, one should be able to invert $\mathbf{\Theta}$ to recover the higher dimensional state, $X$. Another interpretation is that energy sparsity of POD modes automatically guarantees $l_0$ minimization as the sparsity $K$ represents the least number of basis needed to capture the corresponding quantity of energy of the system represented by the training data. From earlier discussions in section 2, we know $P > K$ is a sufficient condition for accurate reconstruction using $l_0$ minimization. Thus, both interpretations require a minimum quantity of sensor data for accurate reconstruction and is verified through numerical experiments in section 4.5. Further, we also correlate the sparsity and coherency metrics with desired reconstruction quality and how these outcomes translate to transient wake flows.

To identify the desired system sparsity $K = N_b$, we define a cumulative energy percentage $\mathbb{E}_K$, which is useful for designing the experiments by choosing the first $K$ number of POD modes as:

$$\mathbb{E}_K = \sum_{k=1}^{K} \frac{\lambda_k}{(\lambda_1 + \lambda_2 + ... + \lambda_M)}, \tag{23}$$

where the eigenvalues $\lambda$ are computed from Equation 9, and $M$ is the total number of positive eigenvalues. As a result, the energy fraction $\mathbb{E}_K$ corresponding to sparsity $K$ for the different $Re$ numbers is shown in Figure 6. For the relatively lower $Re$ number cylinder wake flow $(Re = 100)$, it requires two and five POD modes to capture 95% and 99% of the energy content respectively. The energy-based sparsity $(K)$ for the various $Re$ numbers are summarized in the caption of Figure 6. To relate sparse reconstruction performance across different flow regimes with different values of $K, K_{95}$ we define a normalized system sparsity metric, $K^* = K/K_{95}$ and a normalized sensor sparsity metric, $P^* = P/K_{95}$. This allows us to design an ensemble of numerical experiments in the discretized $P^* - K^*$ space and outcomes can be generalized. In this study, the design space is populated over the range $1 < K^* < 6$ and $1 < P^* < 12$ for all the cases. The lower bound of one is chosen such that the minimally accurate reconstruction captures 95% of the energy or 95% accurate. If one desires a different reconstruction norm, then $K_{95}$ can be changed to $K_{xx}$ without loss of generality and the corresponding $K$-space modified accordingly to modulate the reconstruction quality. Alternately, one can choose $E_K$, the normalized energy fraction metric to represent the desired energy capture as a fraction of $E_{K_{95}}$. As an illustration, the reconstruction is performed for $Re = 100$ with $K_{95} = 2$, by choosing $K$ ranging from 2-12 with an increment of 1, and the corresponding sparse measurements $P$, varied from 2-24 with an increment of 2.

To quantify the $l_2$ reconstruction performance, we define the mean squared error

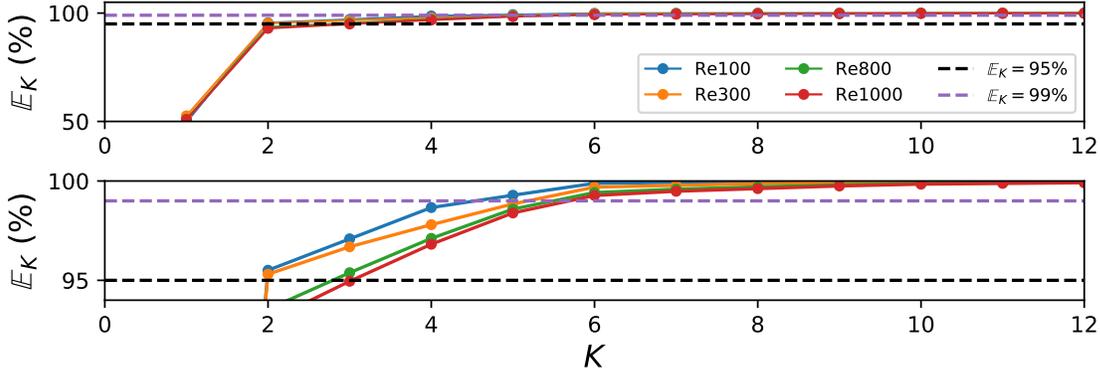

Figure 6: Schematic shows the cumulative energy capture corresponding to various system sparsity levels, $K$, (i.e. the number of POD modes) for cylinder flow at $Re = 100$ to 1000. The bottom figure is a magnified version of the top figure for $\mathbb{E}_K$. For $Re = 100, 300, 800, 1000$: $K_{95} = [2, 2, 3, 4]$ and $K_{99} = [5, 6, 6, 6]$, respectively.

as:

$$\epsilon_{K^*,P^*}^{SCR} = \frac{1}{M} \frac{1}{N} \sum_{j=1}^{M} \sum_{i=1}^{N} (X_{i,j} - \bar{X}_{i,j}^{SCR})^2, \tag{24}$$

where $X$ is the true data, and $\bar{X}^{SCR}$ is the reconstructed field from sparse measurements and $\bar{X}^{POD}$, the reconstructed field from exactly computed POD coefficients. $N$ and $M$ represent the state and snapshot dimensions and related to indices $i$ and $j$, respectively. Similarly, the mean squared error $\epsilon_{K_{95}^*}^{POD}$ and $\epsilon_{K^*}^{POD}$ for the reconstruction from the POD method can be computed as:

$$\epsilon_{K_{95}^*}^{POD} = \frac{1}{M} \frac{1}{N} \sum_{j=1}^{M} \sum_{i=1}^{N} (X_{i,j} - \bar{X}_{i,j}^{POD,K_{95}^*})^2, \tag{25}$$

$$\epsilon_{K^*}^{POD} = \frac{1}{M} \frac{1}{N} \sum_{j=1}^{M} \sum_{i=1}^{N} (X_{i,j} - \bar{X}_{i,j}^{POD,K^*})^2, \tag{26}$$

where $K_{95}*$ is the normalized system sparsity metric (i.e. number of POD modes normalized by $K_{95}$) corresponding to 95% energy capture and $K^* = K/K_{95}$ represents the desired system sparsity. Note that $K_{95}*$ is trivially seen to be unity for this case. Then, the normalized absolute ($\epsilon_1$) and relative ($\epsilon_2$) errors are computed as follows. $\epsilon_1$ represents the reconstruction error normalized by the corresponding POD reconstruction error for 95% energy capture. $\epsilon_2$ represents the normalized error relative to the desired reconstruction accuracy for the chosen system sparsity, $K$. These two error metrics are chosen so as to achieve the twin goals of assessing the overall absolute quality of the reconstruction in a normalized sense ($\epsilon_1$) and if the best possible reconstruction accuracy for the chosen problem set-up (i.e $P^*, K^*$) has been realized. Thus, if the best possible reconstruction for a given $K$ is realized then $\epsilon_2$ will take the same value across different $K^*$. This error metric is used to assess relative dependence of $P^*$ on $K^*$ for a given flow problem in order for this method to be successful. However, $\epsilon_1$ is used to provide an absolute estimate of the reconstruction accuracy and assess minimal values of $P^*, K^*$ needed to recover the chosen flow system.

$$\epsilon_1 = \frac{\epsilon_{K^*,P^*}^{SCR}}{\epsilon_{K_{95}}^{POD}} \ , \ \epsilon_2 = \frac{\epsilon_{K^*,P^*}^{SCR}}{\epsilon_{K^*}^{POD}}. \tag{27}$$

### 4.5. Sparse Reconstruction of Limit-cycle Dynamics in Cylinder Wakes

We performed a series of nearly 1000 POD-based $l_2$ sparse reconstruction experiments corresponding to different points in the $P^* - K^*$ design space and spread over the different $Re$ numbers and sensor placements. In these experiments, the sparse data is obtained form *a priori* high resolution flow field data with randomly placed sparse sensors that change for each snapshot. The focus of this experiment is to mimic a time-varying sensing strategy as is observed in the case of mobile sensors. The random sensor placement in this case is controlled using random seed which is fixed for all the experiments within the design space and for the chosen $Re$.

The contours of $\epsilon_1$ and $\epsilon_2$ for all the experiments at different flow $Re$ numbers, $Re$ designed in the $K^* - P^*$ space are shown in Figure 7. The relative error metric $\epsilon_2$ (the right column in Figure 7), shows that the smaller errors (both light and dark blue regions) are predominantly located over the region where $P^* > K^*$ and separated from the other region using a black diagonal line. For small $P^*$, the normalized error can reach as high as $\mathcal{O}(10^2 - 10^3)$. Since $\epsilon_2$ is normalized by the error contained in the exact $K$-sparse POD reconstruction, this metric represents how effectively does the sparse sensor data approximate the $K$-sparse solution using a $l_2$ framework. In principle, the exact $K$-sparse POD reconstruction is the best possible outcome to expect irrespective of how much sensor data is available as long as $K = N_b$. While this 'best' reconstruction is almost always observed for the higher values of $P^*$ at the different flow regimes and $K^*$ values, there appear to be some exceptions. Notably, for all the different $Re$ numbers considered, a small portion of $\epsilon_2$ in the region abutting the $P^* = K^*$ line shows nearly an order of magnitude higher error, $\mathcal{O}(10^1)$ (colored as green in Figure 7) as compared to the expected values in the range of $\mathcal{O}(1)$. While this trend is observed across different $Re$, the tendency is more predominant at the higher values of flow $Re$ number. This effect can be attributed to 'ineffective' sensor placement, but the dependence of the flow regime is not clear. While addressing the question of what is 'effective' sensor placement [60] is beyond the scope of this work, in this study we simply try out different random number seeds to assess if they provide better SR performance. While this approach is more *ad hoc*, it did yield improvements in reconstruction performance. We observed that there exists sets of random sensor locations that decreased the relative errors from $\mathcal{O}(10^1)$ to $\mathcal{O}(10^0)$ as illustrated in the following investigation. Nonetheless, if the effectiveness of sensor placement remains unknown as is the case in many practical problems, choosing $P^* > 2K^*$ displays a higher probability of accurate reconstruction. In general, $P^* > K^*$ appears to be adequate for effective sparse reconstruction (, i.e., $\epsilon_2 \sim \mathcal{O}(10^0)$) with careful sensor placement.

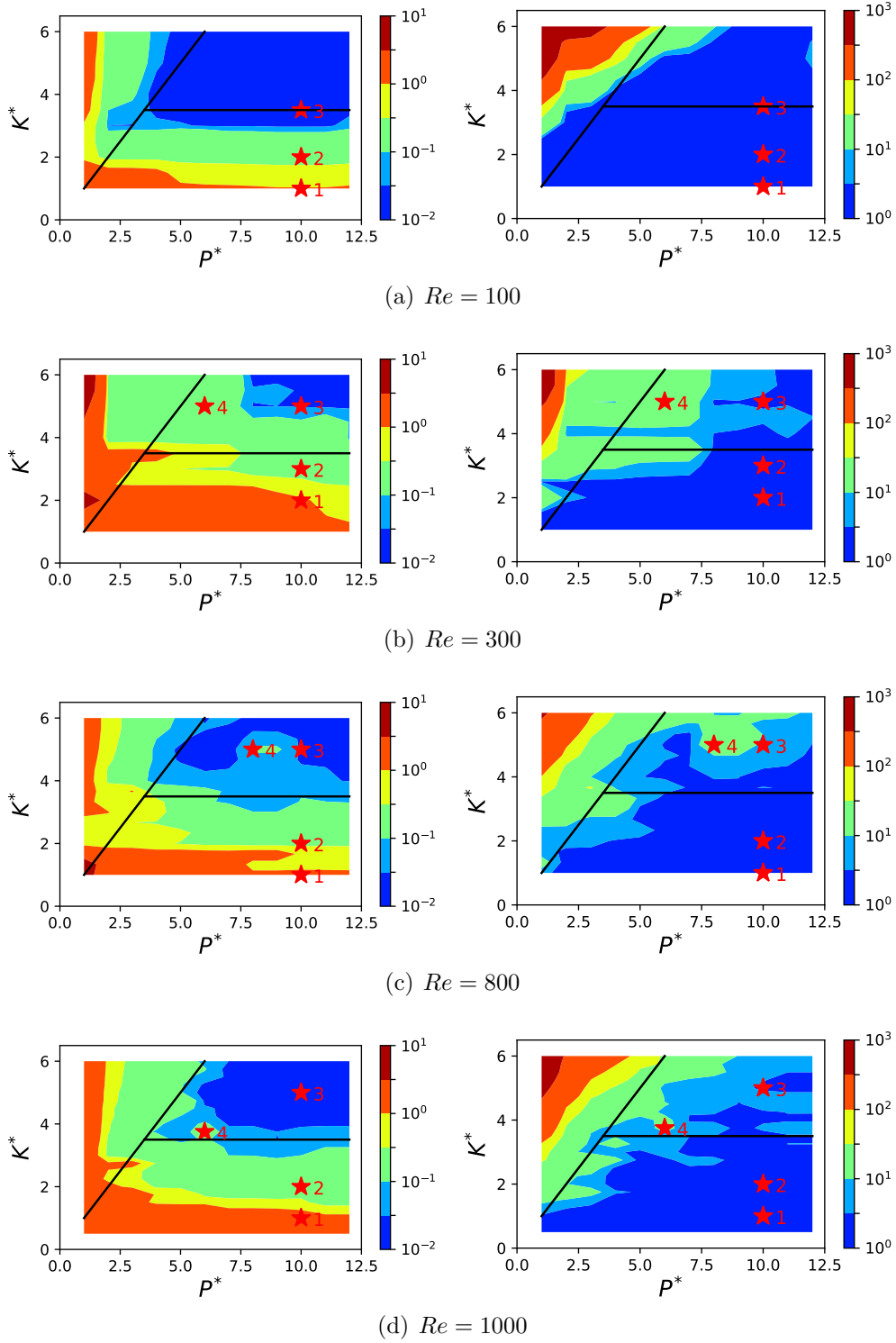

(a) $Re = 100$

(b) $Re = 300$

(c) $Re = 800$

(d) $Re = 1000$

Figure 7: Isocontours of the normalized mean squared $l_2$ POD-based sparse reconstruction errors with respect to the actual (CFD solution) for the designed experiments in the $K^* - P^*$ space for all the $Re$ numbers ($\beta = 101$). Left: normalized absolute error metric, $\epsilon_1$. Right: normalized relative error metric, $\epsilon_2$.

Although the outcomes from the SR experiments are mostly effective for $P^* > K^*$ in terms of the relative error $\epsilon_2$, the absolute error metric $\epsilon_1$ which represents the reconstruction error normalized by a fixed $K$-sparse exact reconstruction containing 95% energy (the left columns in Figure 7) tells a different story. In these series of experiments, the least absolute error $\epsilon_1$ is obtained with high probability when $P^* > K^*$ and $K^* > K^*_{crit}$ and error contours are not so surprisingly $L$-shaped. The choice of $K^*_{crit}$ depends on the the desired reconstruction accuracy. In this study, we chose $\epsilon_1 \approx \mathcal{O}(10^{-1})$, that is, nearly an order of magnitude smaller than the $K_{95}$-sparse exact reconstruction error. This, with a high degree of probability, results in a critical value $K^*_{crit} \approx 3.5$ for the different flow regimes with $Re = 100, 800$ and 1000. For the case with $Re = 300$, the observed critical value for $K^*$ is higher, i.e. $K^*_{crit} \approx 5.0$ possibly due to accumulation of errors from non-optimal sensor placement. We will explore this flow regime in detail in the following discussion. Before that, we summarize the outcomes from our analysis in the following way: in order to obtain accurate sparse reconstructed solutions, one needs in addition to $P^* > K^*$, also $K^* \geq K^*_{crit} \approx 3.5$ for almost all the limit-cycle cylinder wakes considered in this study. The choices of $\epsilon_1 \approx \mathcal{O}(10^{-1})$ and $K^*_{crit} \approx 3.5$ are not arbitrary and chosen after careful investigation of the results. Particularly, we chose three data points as denoted using red stars in Figure 7 and labeled as 1, 2, and 3, corresponding to different levels of error in decreasing order corresponding to $\epsilon_1 \sim \mathcal{O}(10^0)$, $\mathcal{O}(10^{-1})$, and $\mathcal{O}(10^{-2})$, respectively for $Re = 100$, 300, 800, and 1000. We evaluated the corresponding reconstructed solutions using the $l_2$ SCR algorithm for qualitative comparison with the actual data. For conciseness and to minimize repetition, we only compare the line contours between the actual CFD solution and SCR at $T = 0$ for $Re = 100$ in Figure 8 (the top row). It is seen from fig. 8(b) that for the case with $\epsilon_1 \sim \mathcal{O}(10^{-1})$ one observes discrepancies at the wake of the cylinder whereas fig. 8(c) shows that for the case with $\epsilon_1 \sim \mathcal{O}(10^{-2})$ the reconstructed results seem fairly accurate. In addition, the corresponding projected and reconstructed features $a$ (POD coefficients) for the various cases are shown in fig. 8 (the bottom row). The discrepancy in these POD features are small, but still observed for the highest modes with the smallest energy. For $K^* = 1$ in fig. 8a with only two features, such errors that arise from inadequate sensor placement are critical and impact the overall accuracy. Further, although $P^*$ greater than $K^*$ is sufficient to approximate the POD coefficients with appropriate sensor placement, using a small number of modes can only achieve to a limited degree of accuracy in terms of the absolute error metric, $\epsilon_1$. For example, higher numbered modes, such as mode 7 in the case shown in fig. 8c is still of importance, but shows slight inaccuracy. The general rule of thumb to obtain an accurate reconstruction is to retain higher number of modes such as $K^* \geq 3.5$ while satisfying $P^* > K^*$. Similar results for other $Re$ numbers have also been observed as evidenced from figs. 16 , 17, and 18 in the appendix. The corresponding SCR solution (field and coefficients) qualitatively agrees with the actual CFD-generated data. The quantifications at the selected design space points for all the different $Re$ numbers are summarized in table 4.

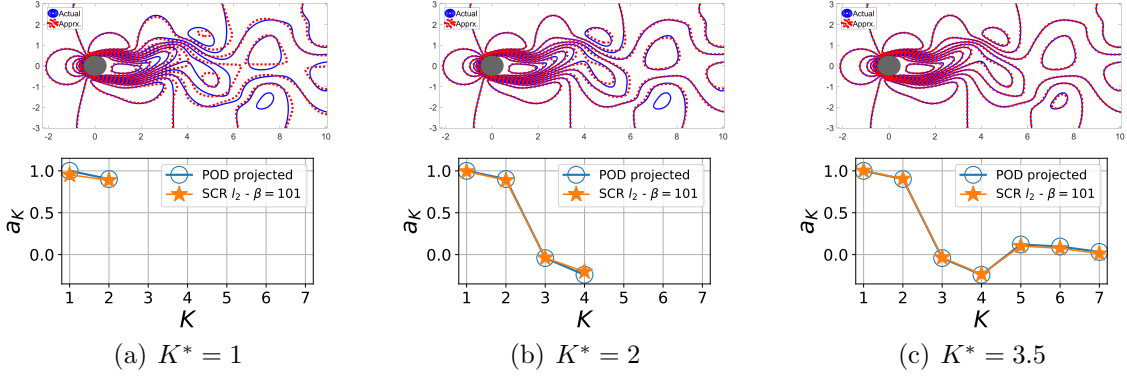

| (a) $K^* = 1$ | (b) $K^* = 2$ | (c) $K^* = 3.5$ |

Figure 8: Top row: we show the line contour comparison of streamwise velocity between the actual CFD solution field (blue) and the energy-based SCR reconstruction (red) for $Re = 100$ ($\beta = 101$) at $P^* = 10$. The contours correspond to a snapshot at time $T = 0.2$. Bottom row: we show the corresponding normalized projected and reconstructed coefficient $a$ from the SCR algorithm. From left to right: $\epsilon_1 = 1.03 \times 10^0$, $3.25 \times 10^{-1}$, $2.85 \times 10^{-2}$ and $\epsilon_2 = 1.03 \times 10^0$, $1.09 \times 10^0$, $1.86 \times 10^0$, respectively.

The SCR performance for regions corresponding to $P^* > K^*$ and $K^* \geq 3.5$ do not always show good accuracy and there exist anamolies near the $P^* = K^*$ boundary. We see this especially for the $Re = 300$ case as shown in Figure 7(b). Nearly half of the region displays surprisingly high errors (green colored error contours) for both $\epsilon_1$ and $\epsilon_2$. A potential cause could be the normalization parameter ($K_{95}$) for both $K$ and $P$. $K_{95}$ corresponds to the integer representation of number of modes to capture 95% of the energy content and $K_{95} = 2$ for both $Re = 100$ and $Re = 300$. This does not reflect the very different length and time-scale content in the wake dynamics at these two flow regimes which skews the estimation of $K^*_{crit}$. Other possible reasons include undesirable errors introduced by non-optimal sensor placement through the matrix, $C$. Such inconsistencies are also observed for the other higher Reynolds numbers figs. 7(c) and 7(d). In our experiment, the random seed parameter ($\beta$) in *Matlab* is used to regulate the random sensor selection. The above results used $\beta = 101$ to identify the random sensor locations for each snapshot. Because this number is randomly chosen rather than using either the physics-based or data-driven approaches, ineffective choice of sensors can result. To explore potential improvements, we have selected additional points labeled as point 4 in Figure 7 for $Re = 300$, 800, and 1000 and assessed the SR accuracy from different choices of $\beta$. For instance, the initial sensor placement from $\beta = 101$ for $Re = 300$ is shown in Figure 9(a). We notice the selected sparse locations biased towards the bottom of the domain region as against the wake region. From our previous studies in [61], locating more sensors around the cylinder and the wake appears to improve the prediction significantly. Consequently, we improved the sensor placement as shown in Figure 9(d) by choosing $\beta = 200$. Comparing fig. 9(b) and 9(e), the reconstructed solutions with the improved sensor selection are more accurate. Although the reconstructed coefficients $a$ show strong visual agreement with the projected $a$ for both values of $\beta$ as shown in Figure 9(c) and 9(f), $\epsilon_1$ and $\epsilon_2$ have decreased by an order of magnitude for $\beta = 200$ as evidenced in Table 4. This indicates that for such sparse fluid dynamics, even small errors in the prediciton of the POD features can impact reconstruction quality which makes sensor placement paramount. Consequently, we apply the improved $\beta$ to simulate the entire $K^* - P^*$ design space for $Re = 300$ and the results are shown in fig. 10. Clearly, both $\epsilon_1$ and $\epsilon_2$ show much reduced errors consistent with observed behavior at other Reynolds numbers in figs. 7(c) and 7(d).

Table 4: Sparse reconstruction performance quantification at selected design points for different $Re$ numbers for both periodic and transient cylinder flows. Symbol "+" indicates that the transient cylinder flow which has temporally evolving behavior. $\epsilon_1$ is the SCR reconstruction error normalized by the exact $K_{95}$-sparse exact POD reconstruction error. $\epsilon_2$ is the SCR reconstruction error normalized by the $K$-sparse exact POD reconstruction error.

| $Re$ | $K_{95}$ | $K$ | $P$ | $K^*$ | $P^*$ | $\mu_u$ | $\mu_v$ | $\beta$ | $\epsilon_1$ | $\epsilon_2$ | Point |
|---|---|---|---|---|---|---|---|---|---|---|---|
| 100 | 2 | 2 | 20 | 1.0 | 10.0 | 1.127 | 1.136 | 101 | 1.03E+00 | 1.03E+00 | 1 |
| | | 4 | 20 | 2.0 | 10.0 | 1.471 | 1.136 | 101 | 3.25E-01 | 1.09E+00 | 2 |
| | | 7 | 20 | 3.5 | 10.0 | 1.471 | 1.136 | 101 | 2.85E-02 | 1.86E+00 | 3 |
| 300 | 2 | 4 | 20 | 2.0 | 10.0 | 1.835 | 3.857 | 101 | 1.07E+00 | 2.28E+00 | 1 |
| | | 6 | 20 | 3.0 | 10.0 | 1.835 | 3.857 | 101 | 1.79E-01 | 2.72E+00 | 2 |
| | | 10 | 20 | 5.0 | 10.0 | 3.050 | 6.050 | 101 | 4.67E-02 | 4.48E+00 | 3 |
| | | 10 | 12 | 5.0 | 6.0 | 2.687 | 4.076 | 101 | 1.57E-01 | 1.50E+01 | 4 |
| | | 10 | 12 | 5.0 | 6.0 | 3.050 | 6.050 | 200 | 5.20E-02 | 4.99E+00 | 4 |
| 800 | 3 | 3 | 30 | 1.0 | 10.0 | 2.657 | 3.093 | 101 | 1.22E+00 | 1.22E+00 | 1 |
| | | 6 | 30 | 2.0 | 10.0 | 3.701 | 4.968 | 101 | 3.83E-01 | 3.04E+00 | 2 |
| | | 15 | 30 | 5.0 | 10.0 | 4.372 | 6.203 | 101 | 4.16E-02 | 6.97E+00 | 3 |
| | | 15 | 24 | 5.0 | 8.0 | 4.372 | 6.264 | 101 | 1.15E-01 | 1.92E+01 | 4 |
| | | 15 | 24 | 5.0 | 8.0 | 6.264 | 5.426 | 132 | 3.62E-02 | 6.07E+00 | 4 |
| 1000 | 4 | 4 | 40 | 1.0 | 10.0 | 2.285 | 3.578 | 101 | 1.20E+00 | 1.20E+00 | 1 |
| | | 8 | 40 | 2.0 | 10.0 | 3.883 | 4.834 | 101 | 2.56E-01 | 2.13E+00 | 2 |
| | | 20 | 40 | 5.0 | 10.0 | 5.467 | 6.592 | 101 | 1.69E-02 | 4.62E+00 | 3 |
| | | 15 | 24 | 3.75 | 6.0 | 3.883 | 4.834 | 101 | 1.45E-01 | 1.26E+01 | 4 |
| | | 15 | 24 | 3.75 | 6.0 | 6.001 | 5.444 | 132 | 8.65E-02 | 7.53E+00 | 4 |
| 100+ | 5 | 22 | 35 | 4.4 | 7.0 | 2.320 | 2.367 | 101 | 1.61E-01 | 1.30E+01 | 5 |
| | | 22 | 35 | 4.4 | 7.0 | 5.010 | 4.953 | 193 | 4.42E-02 | 3.58E+00 | 5 |

Similar improvements in SR accuracy was also observed for design probe 4 at $Re = 800$ and 1000 as shown in Figures 19 and 20 (see appendix) and Table 4. Interestingly, although we barely observe any difference between the actual and reconstructed solution fields for a single snapshot reconstruction for $Re = 800$ in fig. 17 we observe significant reduction in error metrics for the entire ensemble of snapshots as presented in table 4. We relate improved sensor placement to the coherency measures defined in eq. 12. The higher the coherency measure, $\mu$, the more data one needs to reconstruct the fine-scale field using $l_1$ minimization in SCR for the desired reconstruction accuracy. In SCR, it is generally desired to have smaller $\mu$ so that the information can reconstructed with the minimum quantity of data. However, data shown in table 4 for energy-sparse $l_2$-based SCR indicate that $\mu$ mostly increases (although small) with improved sensor placement, i.e., the sensors are more coherent with the POD basis while reducing errors. This may mean that one needs more data to reconstruct accurately using $l_1$ approaches with high probability as per [51]. However, for data-driven sparse POD basis containing information about physically relevant flow structures, increasing coherence by placing sensors in alignment with the flow topology helps improve reconstruction accuracy for a fixed sensor budget. That said, $\mu$ is still a useful metric to ensure all the different basis are sufficiently excited by the data when one uses a large candidate basis space.

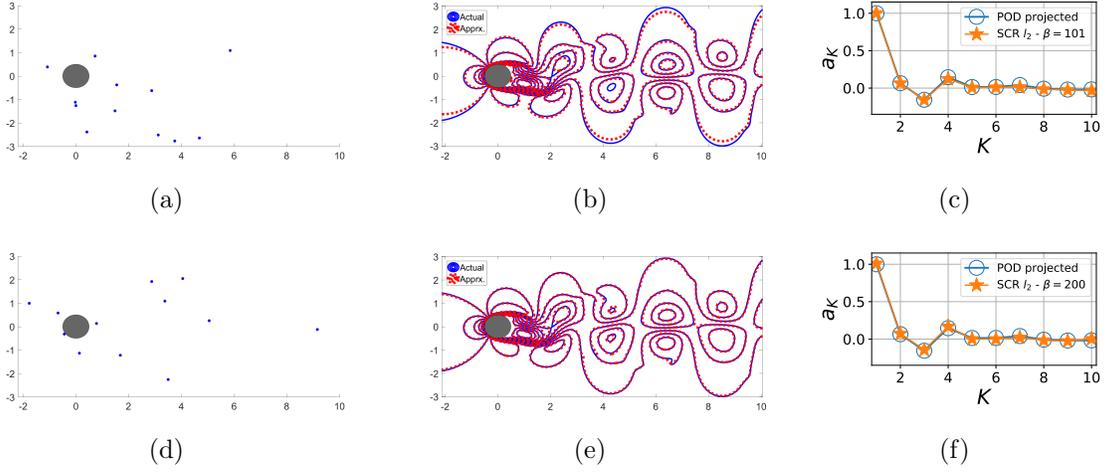

Figure 9: The sensor locations (left column), the true and reconstructed solutions (middle column), and the normalized projected and reconstructed coefficient $a$ from the POD-SCR algorithm (right column) for $Re = 300$ from $\beta = 101$ (top row) and 200 (bottom row). The snapshot chosen for this representation corresponds to $T = 0.2$. Top: $\epsilon_1$=1.57E-01 and $\epsilon_2$=1.50E+01. Bottom: $\epsilon_1$=5.20E-02 and $\epsilon_2$=4.99E+00.

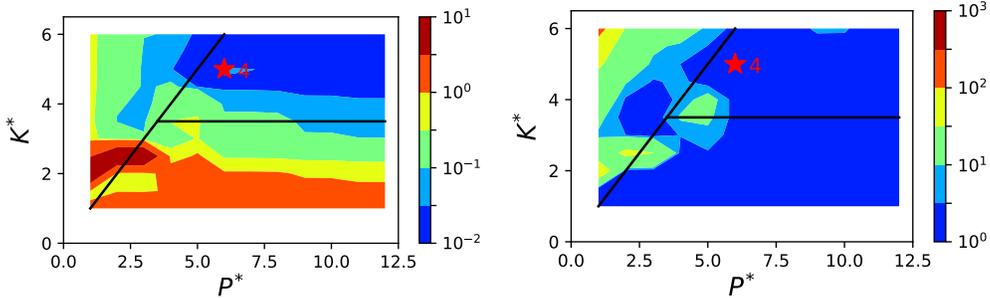

Figure 10: Isocontours of the normalized mean squared error metrics for POD-based sparse reconstruction with respect to the actual (CFD solution) for the periodic cylinder flow at $Re = 300$ with improved sensor placement ($\beta = 200$). Left: normalized absolute error metric, $\epsilon_1$. Right: normalized relative error metric, $\epsilon_2$.

### 4.6. Sparse Reconstruction of Transient Cylinder Wakes

The previous section focused on examining the SR performance of limit-cycle cylinder wake dynamics using POD basis from similar limit-cycle dynamics over a range of $Re$ numbers. For this baseline case the study showed that the POD-SCR algorithm performs reasonably well if certain conditions are satisfied. However, their extension to transient cylinder wakes where the nonlinear dynamics involving instability growth with transition of a steady wake into vortex shedding limit-cycle system is not guaranteed. In this section, we explore and assess the performance of the POD-SCR algorithm for such temporally evolving dynamics at $Re = 100$. For this case, we choose 340 snapshots of data for $Re = 100$ which corresponds to a non-dimensional time unit $T = 68$ with uniform temporal spacing of $dT = 0.2s$. The first half of the data lies in the transient or temporally growing unstable regime whereas the second half represents a stable limit-cycle dynamical system. The temporal evolution of the first three POD coefficients are shown in Figure 14. For this data the characteristic dimensionality, $K_{95} = 5$ and POD-SCR is performed over a $P^* - K^*$ design space where the sparsity $K$ is varied from $4 - 30$ with increments of 2 and the sparse sensors $P$ are varied over $5 - 60$ with increments of 5. This discretization of the design space is intentionally matched with that for limit-cycle analysis in the earlier section. As before, we perform reconstruction using random sensor placement with a seed of $\beta = 101$. The normalized error metrics for this SCR (both absolute and relative) are shown in fig. 12(a). As was observed for the limit cycle case, we broadly see that the relative normalized error metric, $\epsilon_2$ to be $\mathcal{O}(1)$ for $P^* > K^*$. However, there exists a consistent region of high error above $K^* \gtrapprox 3$.

As before, we probe this anomaly in the error field by choosing a point in the appropriate high error region of the design space identified by the number 5 which corresponds to $\epsilon_1 \sim \mathcal{O}(10^{-1})$ and $\epsilon_2 \sim \mathcal{O}(10^1)$. As seen from Figure 13(a),the random sensors with seed $\beta = 101$ are located densely around the cylinder and above the wake with only few measurement points within the wake. While this may be reasonable when the wake is steady, it is certainly not sufficient for use in a case where the wake becomes unsteady with vortex shedding. For this sensor arrangement, we also compare the line contours of the streamwise velocity field for the actual CFD solution and outcomes from the POD-SCR algorithm. Figure 13(a) compares the actual CFD generated snapshot ($T = 68$) with the reconstructed solution which should be very close to exact $K$-sparse reconstruction for $K^* = 4.4$. Not surprisingly, the reconstructed field in the wake

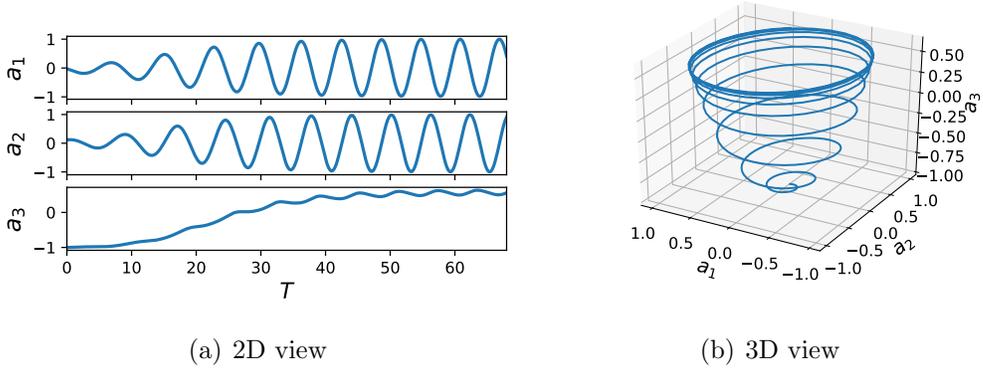

(a) 2D view

(b) 3D view

Figure 11: The temporal evolution of the first three normalized POD coefficients for the transient cylinder flow at $Re = 100$.

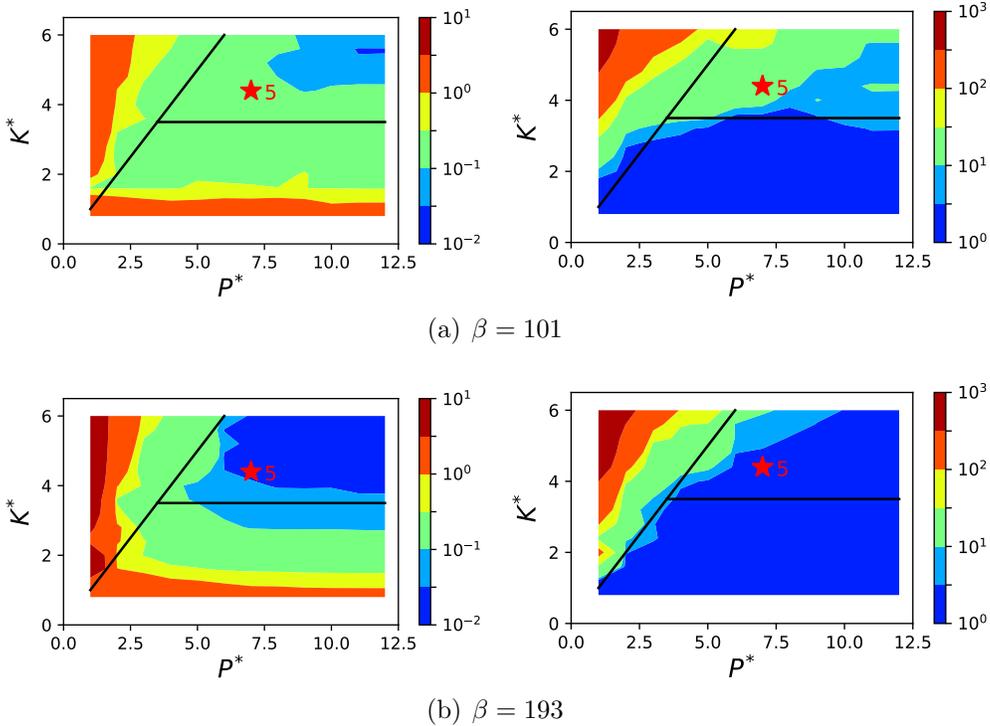

(a) $\beta = 101$

(b) $\beta = 193$

Figure 12: Isocontours of the normalized mean squared $l_2$ POD-based sparse reconstruction errors with respect to the actual (CFD solution) for the transient cylinder flow at $Re = 100$ with different seedings. Left: normalized absolute error metric, $\epsilon_1$. Right: normalized relative error metric, $\epsilon_2$.

region shows perceptible disparity due to paucity of sensors. We explore improved sensor placement with $\beta = 193$ as shown in Figure 13(b). While this approach may sound *ad hoc*, we adopted this strategy as a substitute for data-driven optimal sensor placement algorithms [60] which will be explored in future studies. This particular

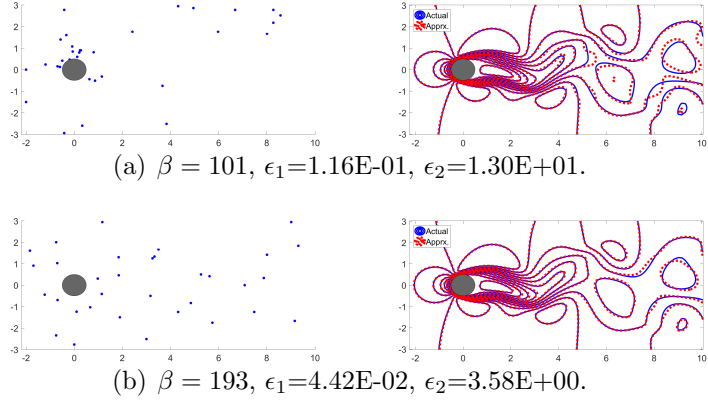

(a) $\beta = 101$, $\epsilon_1$=1.16E-01, $\epsilon_2$=1.30E+01.

(b) $\beta = 193$, $\epsilon_1$=4.42E-02, $\epsilon_2$=3.58E+00.

Figure 13: Comparison of line contours of streamwise velocity at time $T = 68$ for the actual CFD solution field (blue) and the POD-SCR reconstruction (red) from $\beta = 101$ and 193 for the transient cylinder flow at $Re = 100$. $K^* = 4.4$, $P^* = 7$.

selection of random locations place the sensors more evenly across the wake. As a result, the SCR solution for $\beta = 193$ is vastly improved with both $\epsilon_1$ and $\epsilon_2$ having decreased by nearly an order of magnitude as compared to the earlier set-up. To verify the consistency of this trend, we apply this improved sensor placement to reconstruct the entire $P^* - K^*$ design space. The resulting error field shown in fig. 12(b), display $\mathcal{O}(1)$ values for $\epsilon_2$ predominantly in the region $P^* \geq K^*$. In addition, the contours of $\epsilon_1$ show $\mathcal{O}(10^{-1})$ values for $K^* \geq 3.5$ which is consistent with the observations for the limit-cycle cylinder wake dynamics. The reconstructed features $a$ for the two different sensor placement choices corresponding to $\beta = 101 \& 193$ are shown in Figure 14 for a single snapshot corresponding to $T = 68$. The visible errors are observed from the third feature onwards which albeit small, impact the full field reconstruction significantly. In summary, while the magnitude of the POD-SCR error metrics for the transient cylinder wake are comparable to that of the limit-cycle wake dynamics,they are much more sensitive to the sensor placement due to the evolving nature of the flow.

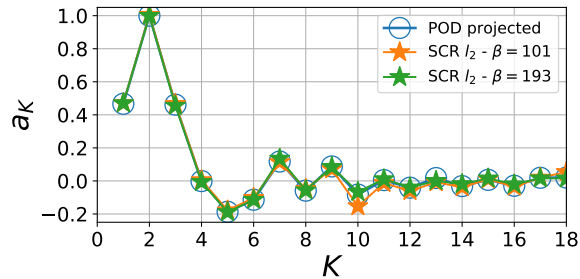

Figure 14: The normalized projected and reconstructed coefficient $a$ for the snapshot at $T = 68$ from the POD-SCR algorithm with $\beta = 101$ and 193 for the transient cylinder flow at $Re = 100$.

## 5. Conclusion

In this article, we explore the interplay of sensor quantity, placement and system sparsity using optimal data-driven basis on reconstruction performance using a efficient sparse POD reconstruction algorithm, also known as gappy POD, that bypasses construction of the full measurement matrix. In general, the choice of basis used in SCR plays a crucial role in the overall performance as it determines the quantity of sensors and their placement for a desired recovery quality. Employing POD/SVD basis also allows for efficient energy-based SR as long as they span the sparse data, i.e. the most energetic POD basis for the training data are also the energetic structures in the data to be reconstructed. Unlike generic basis spaces such as Fourier or radial basis functions, the data-driven basis needs to be flow relevant as retaining the $K$ most energetic modes for reconstruction is also the $K$-sparse solution for the given sensor locations. This is in contrast to common sparse reconstruction algorithms in image processing that are unaware of the basis hierarchy for a given data *a priori* and require searching using sparsity promoting $l_1$ minimization methods. Such energy hierarchy aware sparse reconstruction using $l_2$ techniques also implicitly satisfies $l_0$ minimization.

A baseline case that satisfies the flow relevance condition is the SR of limit-cycle cylinder wake flow dynamics using POD basis learnt from similar data. From our analysis of numerical experiments, we show that the $K-$most energetic basis are the same ones retained from a sparsity promoting $l_1$ minimization reconstruction. This established, we further show that the condition for accurate sparse reconstruction (based on relative error metrics, $\epsilon_2 = \mathcal{O}(10^0)$) using this algorithm is achieved for combinations of the desired system sparsity, $K^*$ and sensor sparsity, $P^*$ such that $P^* \gtrapprox K^*$. Simply put, this states that one needs more sensors than the desired dimension of reconstruction in the chosen basis space which is a key underlying principle of SCR methods. While this result is independent of the flow Reynolds numbers ($Re = 100, 300, 800$ and $1000$), there exists some points in the $P^* - K^*$ design space that show nearly an order of magnitude higher error for $P^* \gtrapprox K^*$ as shown in fig. 7(b). Such anomalies are more prominent at higher Reynolds and are related to the random sensor placement that often tend to be inadequate for the desired reconstruction quality. The connection between erroneous sensor placement and higher Reynolds numbers can be understood as follows. At higher Reynolds numbers, the flow has more (POD) features that need to included in the correct proportions to recreate the true field. These multiple features tend to be more difficult to capture using very few sensors and explains the sensitive predictions. It turns out that using a criterion of $P^* \gtrapprox 2K^*$ is more conservative and produces more robust results. To ascertain the connection between sensor placement and the anomalous prediction errors, we explored alternative sensor locations that coincide with the regions of strong wake flow dynamics. This strategy significantly improved the SR error metrics across the different regimes due to improved estimations of the POD features and the reconstructed fields. While the relative normalized error metric help compare across different choices of $K^*$, the more relevant error metric is the absolute error ($\epsilon_1$) normalized by the $K_{95}$- sparse exact reconstruction error, $\epsilon_{K_{95}}^{POD}$. In this work, we observe that $\epsilon_1 = O(10^{-1})$, i.e. an order of magnitude smaller than $\epsilon_{K_{95}}^{POD}$ offers reasonably accurate reconstruction and is achieved with high probability when $P^* > K^*$ and $K^* \gtrapprox K_{crit}^* \approx 3.5$ as long as the sensor placement is satisfactory. Although small, improved sensor placement led to increase in coherency metric indicating enhanced coherence with the POD basis while reducing reconstruction errors. This is not surprising given that the sensors were configured to align with the flow coherent structures which are also captured in the most energetic POD basis and hence, increase $\mu$. Larger $\mu$ may imply more sensors are needed to achieve the desired reconstruction quality in $l_1$-based methods, but is not the case for this $l_2$-based energy sparse POD-SCR framework.

We followed the earlier analysis for a limit-cycle system with a flow dynamics involving multiple regimes. We applied the POD-SCR framework to snapshots of data representing the transient evolution of the cylinder wake from the point of onset of the instability to its growth into a stable limit cycle system as shown in fig. 14. For this case the observed errors metrics and the corresponding sparsity metric bounds for the chosen reconstruction accuracy, i.e. $\epsilon_1 = O(10^{-1})$ and $\epsilon_2 = O(10^0)$ are comparable to that for the limit-cycle system. Particularly, we note that $P^* \gtrapprox K^*$ and $K^* \gtrapprox K_{crit}^* \approx 3.5$ produce desirable performance with a high degree of probability for random sensor placement in spite of the time-dependent flow regimes. However, these numerical experiments displayed stronger sensitivity to sensor placement as the evolving dynamics require sensors to be placed at different locations to capture the different phenomena. In the future, we intend to explore situations where the training and prediction regimes have dynamics that differ significantly to assess the extent of applicability of such data-driven basis enabled sparse reconstruction methods. A plausible approach would be to build a library of basis that represent the different flow regimes and use $l_1$ sparse reconstruction ideas from compressive sensing for sparse recovery for the flow fields.

## 6. Acknowledgments

We acknowledge computational resources from HPCC and start-up research funds from the Oklahoma State University.

## 7. Author Contributions

BJ conceptualized the research with input from CL. CL developed the sparse reconstruction codes with input from BJ. BJ and CL analyzed the data. CL developed the first draft of the manuscript and BJ produced the final written manuscript.

## 8. Conflicts of Interest

The authors declare no conflict of interest.

## 9. Appendix

### 9.1. POD Modes for Cylinder Wake Limit-cycle Dynamics

Shown below are the most energetic POD basis structures visualized as isocontours for the four different flow regimes considered. The most energetic modes are ordered from left to right. The two most energetic modes clearly show the characteristic vortex shedding structures in the wake with their characteristic sizes decreasing with increasing inertial dynamics. One would see more flow relevant structures in the higher mdoes (with lower energy) for the higher Reynolds numbers.

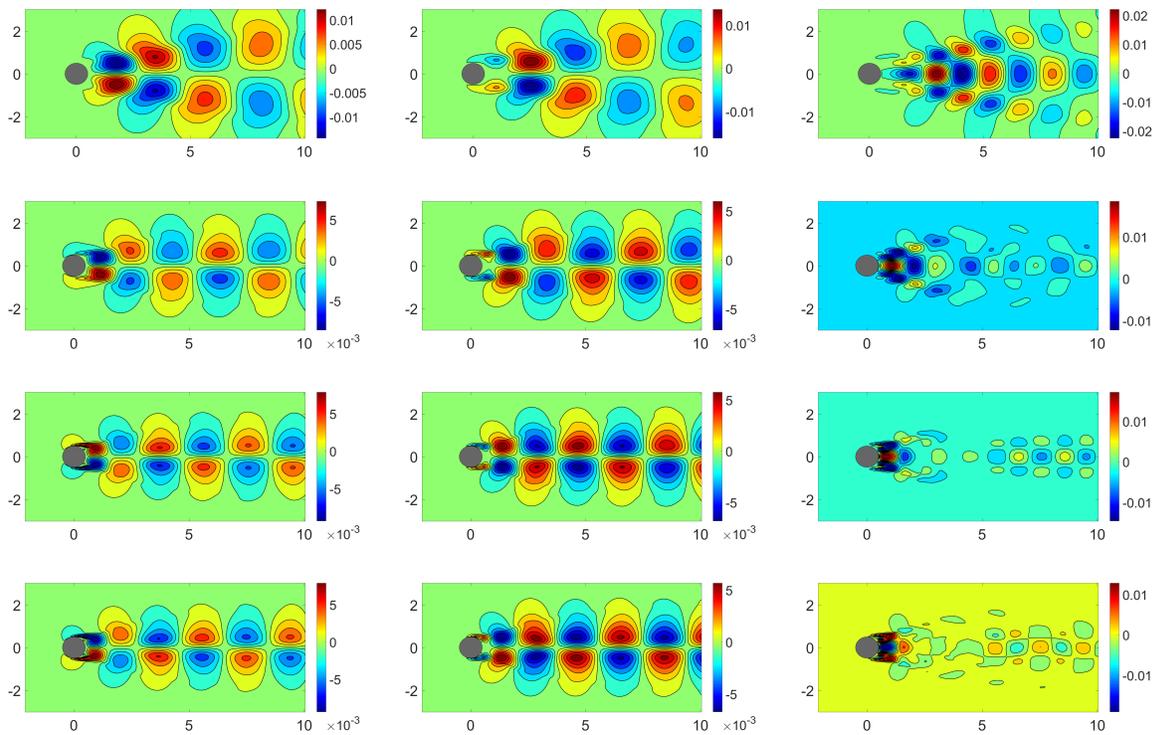

Figure 15: Isocontours of the three most energetic modes (from left to right) for the cylinder flow at different $Re$ number. From top to bottom:$Re = 100$, $300$, $800$, and $1000$.

### 9.2. Extended Analysis of Wake Reconstruction Performance at Higher Reynolds Numbers

The schematics shown in this section illustrate the detailed analysis of the reconstruction performance at the selected deisgn space points 1, 2&3 corresponding to different values of SR errors. These are shown in addition to the $Re = 100$ case discussed in the main sections of the article. This additional analysis reinforce the outcomes discussed in the paper.

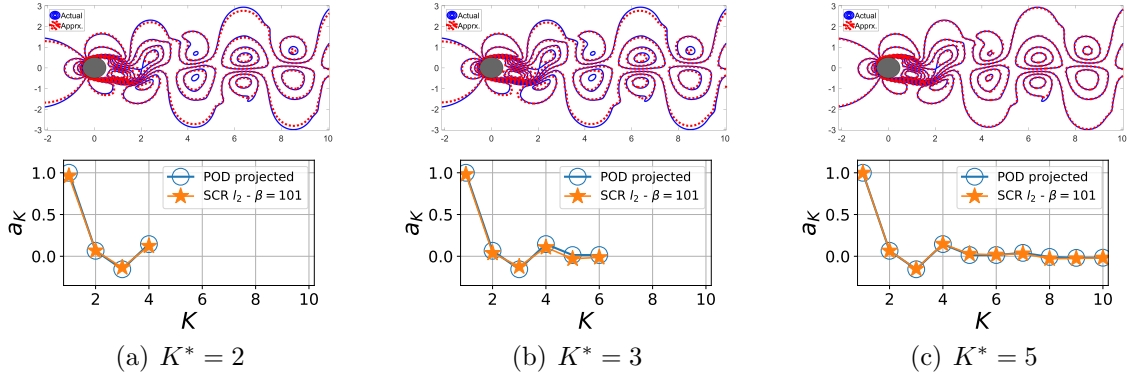

(a) $K^* = 2$        (b) $K^* = 3$        (c) $K^* = 5$

Figure 16: Top row: the line contour comparison of streamwise velocity for the actual CFD solution field (blue) and the GPOD reconstruction (red) for $Re = 300$ ($\beta = 101$) at $P^* = 10$. The snapshot used for this representation corresponds to $T = 0.2$. Bottom row: the corresponding normalized projected and reconstructed coefficient $a$ from the GPOD algorithm. From left to right: $\epsilon_1$=1.07E+00, 1.79E-01, 4.67E-02 and $\epsilon_2$=2.28E+00, 2.72E+00, 4.48E+00, respectively.

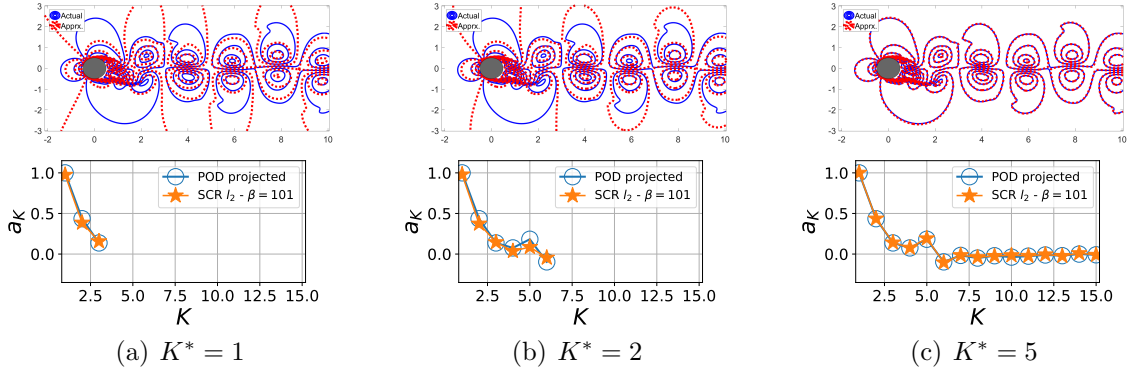

(a) $K^* = 1$        (b) $K^* = 2$        (c) $K^* = 5$

Figure 17: Top row: the line contour comparison of streamwise velocity for the actual CFD solution field (blue) and the POD SCR (red) for $Re = 800$ ($\beta = 101$) at $P^* = 10$. The snapshot used for this representation corresponds to $T = 0.2$. Bottom row: the corresponding normalized projected and reconstructed coefficient $a$ from the GPOD algorithm. From left to right: $\epsilon_1$=1.22E+00, 3.83E-01, 4.16E-02 and $\epsilon_2$=1.22E+00, 3.04E+00, 6.97E+00, respectively.

### 9.3. Effect of Improved Sensor Placement for $Re = 800$ and $Re = 1000$

The schematics shown in this section illustrate the effect of improved sensor placement for higher Reynolds numbers. These are shown as additional material to the $Re = 100$ case discussed in the main sections of the article. These additional cases reinforce the outcomes discussed in the paper.

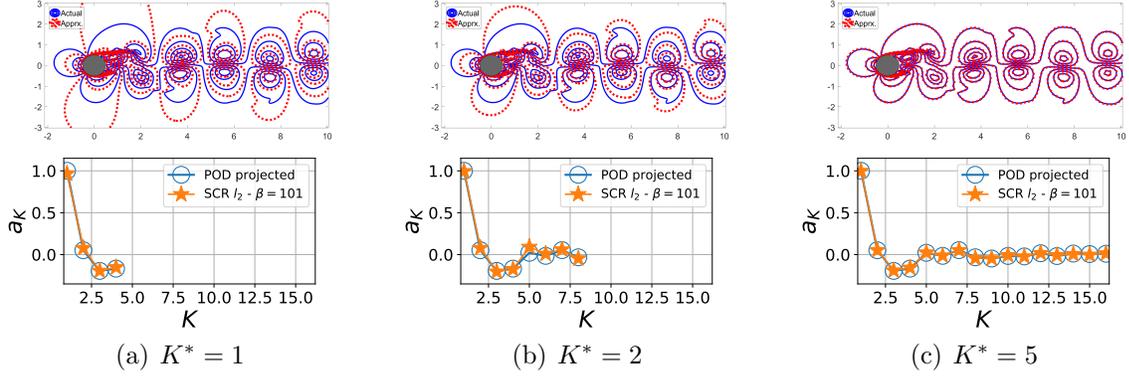

(a) $K^* = 1$    (b) $K^* = 2$    (c) $K^* = 5$

Figure 18: Top row: the line contour comparison of streamwise velocity for the actual CFD solution field (blue) and the POD SCR algorithm (red) for $Re = 1000$ ($\beta = 101$) at $P^* = 10$. The snapshot used for this representation corresponds to $T = 0.2$. Bottom row: the corresponding normalized projected and reconstructed coefficient $a$ from the GPOD algorithm. From left to right: $\epsilon_1 = 1.20\text{E}+00$, 2.56E-01, 1.69E-02 and $\epsilon_2 = 1.20\text{E}+00$, 2.13E+00, 4.62E+00, respectively.

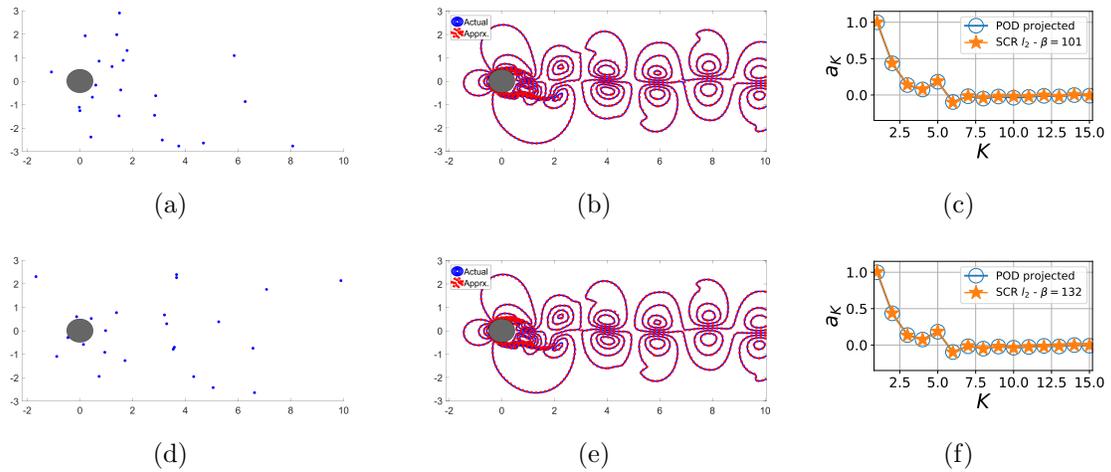

(a)    (b)    (c)

(d)    (e)    (f)

Figure 19: The sensor placements (left column), the true and reconstructed solutions (middle column), and the normalized projected and reconstructed coefficient $a$ from the POD-SCR algorithm (right column) for $Re = 800$ with $\beta = 101$ (top) and 132 (bottom). The snapshot used for this representation corresponds to $T = 0.2$. Top: $\epsilon_1 = 1.15\text{E}-01$ and $\epsilon_2 = 1.92\text{E}+01$. Bottom: $\epsilon_1 = 3.62\text{E}-02$ and $\epsilon_2 = 6.07\text{E}+00$.

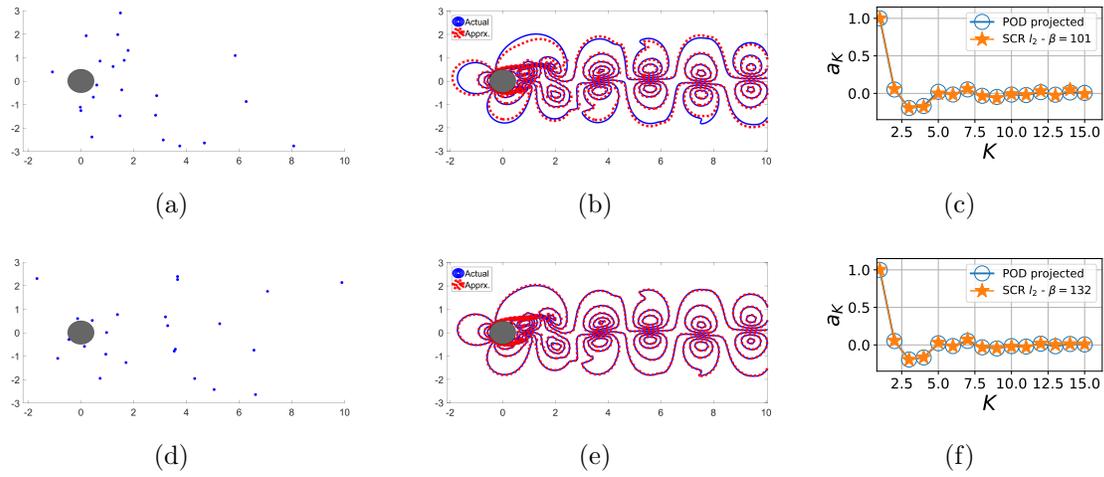

Figure 20: The sensor placements (left column), the true and reconstructed solutions (middle column), and the normalized projected and reconstructed coefficient $a$ from the POD-SCR algorithm (right column) for $Re = 1000$ with $\beta = 101$ (top) and 132 (bottom). The snapshot used for this representation corresponds to $T = 0.2$. Top: $\epsilon_1 = 1.45\text{E-}01$ and $\epsilon_2 = 1.26\text{E+}01$. Bottom: $\epsilon_1 = 8.65\text{E-}02$ and $\epsilon_2 = 7.53\text{E+}00$.